\documentclass[times,twocolumn,final]{elsarticle}

\usepackage{medima}
\usepackage{framed,multirow}

\usepackage{amssymb}
\usepackage{latexsym}
\usepackage{graphicx}
\usepackage{makecell}
\usepackage{array}
\usepackage{mdwmath}
\usepackage{mdwtab}
\usepackage{eqparbox}
\usepackage{url}
\usepackage{multirow}
\usepackage{algorithmic}
\usepackage{algorithm}
\usepackage{url}
\usepackage{xcolor}
\usepackage{multirow}
\usepackage{hyperref}

\definecolor{newcolor}{rgb}{.8,.349,.1}

\journal{Medical Image Analysis}

\begin{document}

\verso{Yichi Zhang \textit{et~al.}}

\begin{frontmatter}

\title{SemiSAM+: Rethinking Semi-Supervised Medical Image Segmentation in the Era of Foundation Models}%

\author[1,2]{Yichi \snm{Zhang}}
\author[1]{Bohao \snm{Lv}}
\author[2,3]{Le \snm{Xue}}
\author[2,4]{Wenbo \snm{Zhang}}
\author[1,2]{Yuchen \snm{Liu}}
\author[5]{Yu \snm{Fu}}
\author[1,2]{Yuan \snm{Cheng}}
\author[1,2]{Yuan \snm{Qi}}

\address[1]{Artificial Intelligence Innovation and Incubation Institute, Fudan University, Shanghai, China.}
\address[2]{Shanghai Academy of Artificial Intelligence for Science, Shanghai, China.}
\address[3]{Huashan Hospital, Fudan University, Shanghai, China.}
\address[4]{Human Phenome Institute, Fudan University, Shanghai, China.}
\address[5]{School of Information Science and Engineering, Lanzhou University, Lanzhou, China}

\received{}
\finalform{}
\accepted{}
\availableonline{}

\begin{abstract}
Deep learning-based medical image segmentation typically requires large amount of labeled data for training, making it less applicable in clinical settings due to high annotation cost. Semi-supervised learning (SSL) has emerged as an appealing strategy due to its less dependence on acquiring abundant annotations from experts compared to fully supervised methods.
Beyond existing model-centric advancements of SSL by designing novel regularization strategies, we anticipate a paradigmatic shift due to the emergence of promptable segmentation foundation models with universal segmentation capabilities using positional prompts represented by Segment Anything Model (SAM).
In this paper, we present \textbf{SemiSAM+}, a foundation model-driven SSL framework to efficiently learn from limited labeled data for medical image segmentation.
SemiSAM+ consists of one or multiple promptable foundation models as \textbf{generalist models}, and a trainable task-specific segmentation model as \textbf{specialist model}.
For a given new segmentation task, the training is based on the specialist-generalist collaborative learning procedure, where the trainable specialist model delivers positional prompts to interact with the frozen generalist models to acquire pseudo-labels, and then the generalist model output provides the specialist model with informative and efficient supervision which benefits the automatic segmentation and prompt generation in turn.
Extensive experiments on two public datasets and one in-house clinical dataset demonstrate that SemiSAM+ achieves significant performance improvement, especially under extremely limited annotation scenarios, and shows strong efficiency as a plug-and-play strategy that can be easily adapted to different specialist and generalist models.
We hope this work will act as catalysts to promote future in-depth research in annotation-efficient medical image segmentation in the era of foundation models, ultimately benefiting clinical-applicable healthcare practices.

\end{abstract}


\end{frontmatter}



\section{Introduction}

\begin{figure*}[t]
	\includegraphics[width=18cm]{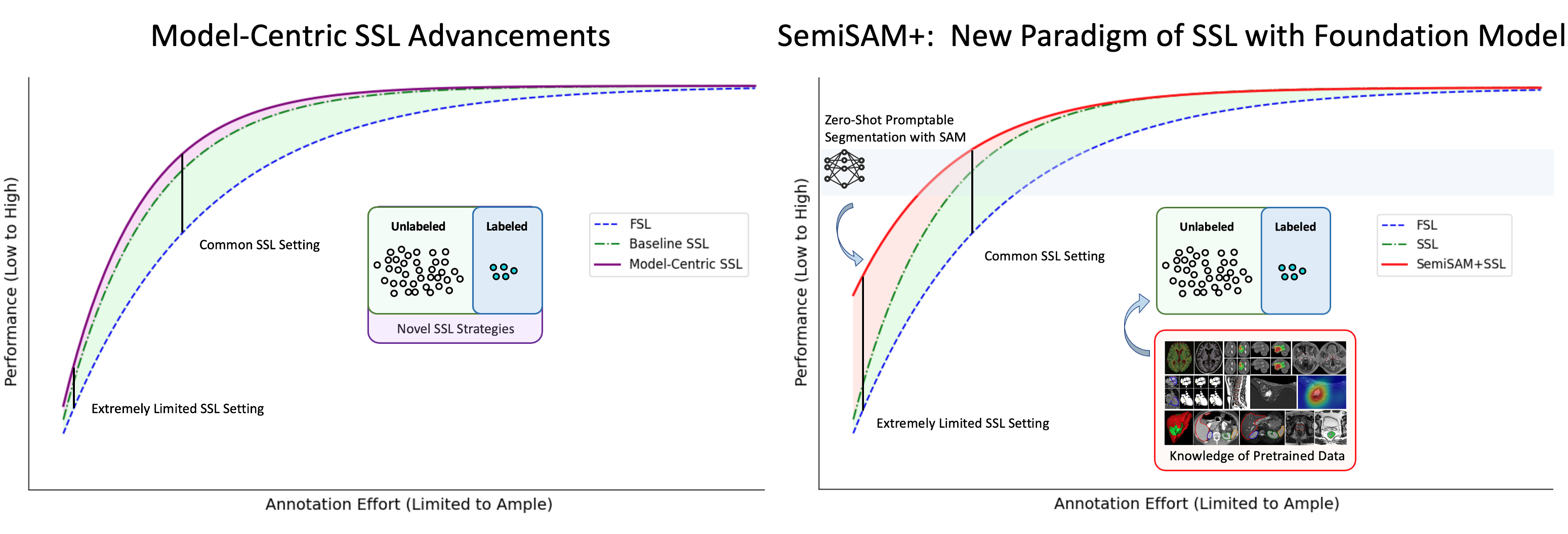}
	\caption{Semi-supervised learning aims to utilize unlabeled data in conjunction with limited amount of labeled data to improve the performance (from blue to green). Existing model-centric advancements of SSL aims to exploit more efficient utilization of unlabeled data for better performance (from green to purple). SemiSAM+ represents a new paradigm to exploit pre-trained knowledge of foundation model to assist in SSL (from green to red).}w
	\label{highlight}
\end{figure*}

Medical image segmentation aims to delineate specific anatomical structures and pathological regions from medical images and serves as a fundamental component in many image-guided therapies. Accurate segmentation provides essential volumetric and morphological information of organs and tumors that facilitates various clinical applications, such as monitoring disease progression, treatment planning, and guiding therapeutic procedures \citep{MSD,lalande2021deep,AbdomenCT-1K,zhang2024nasalseg}. Although deep learning has shown remarkable success in medical image segmentation, the dependence on large-scale annotated datasets for training remains a significant bottleneck, particularly in medical domains where expert annotations are costly and time-consuming \citep{tajbakhsh2020embracing,shi2024beyond}.
The challenge of obtaining extensive annotated medical datasets stems from several factors. First, medical image annotation requires domain expertise from clinical professionals, making the labeling process expensive and resource-intensive. Second, several commonly used medical imaging modalities, such as CT and MRI are 3D volumetric data, requiring slice-by-slice annotation that substantially increases the manual workload compared to 2D image annotation. Additionally, data sharing restrictions between medical institutions further limit the availability of annotated datasets \citep{dumont2021overcoming}.

To address the challenge of manual annotation burden, researchers have made substantial efforts in developing annotation-efficient deep learning methods for medical image segmentation \citep{cheplygina2019not,zhang2021exploiting,zhao2023one}. 
Since unlabeled data are often abundant in practice, semi-supervised learning (SSL) presents an attractive solution and promising direction by effectively using readily available unlabeled data alongside a limited set of labeled data \citep{SemiSurvey}.
Different from generating pseudo-labels and updating the segmentation model in an iterative manner \citep{lee2013pseudo}, recent progress in semi-supervised medical image segmentation has focused on incorporating unlabeled data into the training procedure with unsupervised regularization, such as enforcing the consistency under different data perturbations \citep{yu2019uncertainty,zhang2023uncertainty} or different models \citep{luo2022semi,song2024sdcl}.
In conclusion, these works predominantly focus on model-centric advancements of SSL in the design of novel strategies for generating unsupervised supervision signals to seek more efficient utilization of dark knowledge inside the unlabeled images.
When applying to new segmentation tasks, these semi-supervised methods still appreciate optimistic budgets to annotate a small subset to obtain acceptable performance, struggling to achieve satisfactory outcomes when faced scenarios with extremely limited annotations.

\begin{figure*}[t]
        \centering
	\includegraphics[width=18cm]{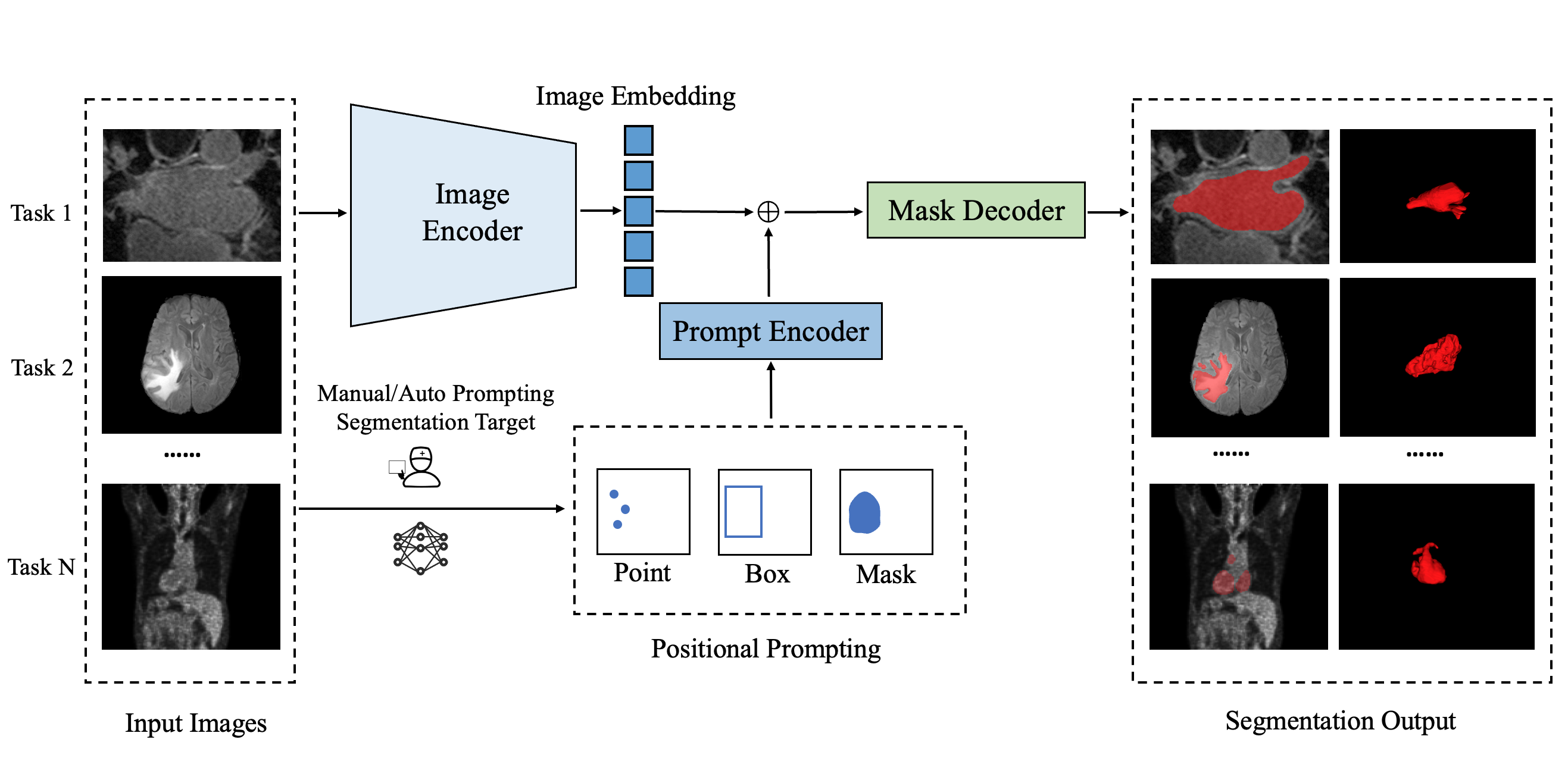}
	\caption{An overview of segmentation foundation model represented by Segment Anything Model (SAM), which adopts an image encoder to extract image embeddings, a prompt encoder to integrate user interactions via different prompt modes, and a mask decoder to predict segmentation masks by fusing image embeddings and prompt embeddings. For any given dataset, the model can segment any target out of the image based on the positional prompting, which demonstrate universal segmentation capability to any new tasks.}
	\label{SAM}
\end{figure*}

Recent years have witnessed the emergence of foundation models in computer vision \citep{awais2023foundational}, notably represented by the Segment Anything Model (SAM) \citep{SAM} for image segmentation, which demonstrates impressive zero-shot generalization capabilities across various natural image segmentation tasks \citep{semanticSAM,trackanything}. Although recent studies have revealed SAM’s limited performance in medical image segmentation due to the differences between natural images and medical images \citep{SAM4MIS,SAM2-MIS}, it still presents new opportunities for addressing the annotation scarcity serving as a reliable pseudo-label generator to guide the segmentation task when manually annotated images are scarce by leveraging the knowledge base of foundation models.

As an initial attempt to exploit SAM-like promptable foundation models for semi-supervised medical image segmentation, our previous work SemiSAM \citep{zhang2024semisam} aims to exploit the potential of SAM by adding an additional supervision branch for consistency regularization, which has been implemented and extended by several researches \citep{ali2024review} to demonstrate the effectiveness. As illustrated in Fig. \ref{highlight}, existing model-centric SSL advancements aims to design novel regularization strategies for more efficient utilization of unlabeled data for training. 
In common SSL settings like using 10\% or 20\% labeled data, these methods can achieve satisfactory performance improvements. However, when confronted with scenarios characterized by extremely scarce annotations like using only one or a few labeled images, they struggle to achieve desirable outcomes.
One possible reason is that the model cannot learn sufficient discriminative information with extremely limited supervision. As a result, the predictions are full of randomness and the improvement of exploiting unlabeled data is limited.
Instead of this paradigm, we aim to utilize pre-trained knowledge of foundation model to guide the learning procedure, which demonstrate significant improvement especially in extremely limited annotation scenario.

In this paper, we present \textbf{SemiSAM+}, a foundation model-driven SSL framework to efficiently learn from limited labeled data for medical image segmentation.
SemiSAM+ contains a trainable task-specific segmentation model as \textbf{specialist model} for downstream segmentation task and one or multiple promptable foundation models as \textbf{generalist models} pre-trained on large-scale datasets with zero-shot generalization ability.
Following existing SSL frameworks, the specialist model is updated using both supervised loss based on labeled data and unsupervised regularization based on unlabeled data. 
In SemiSAM+, an additional confidence-aware regularization is adapted for effective utilization of generalist model's supervision to avoid possible misguidance when encountering extremely limited annotation.
During training, the trainable specialist model delivers positional prompts to interact with the frozen generalist model to acquire pseudo-labels. Then the generalist model output provides the specialist model with informative and efficient supervision which benefits the automatic segmentation and prompt generation in turn.

As an extension of the preliminary version \citep{zhang2024semisam}, SemiSAM+ incorporates substantial methodological advancements with extensive experiments to show strong efficiency as a plug-and-play method applicable to different specialist and generalist models. Additionally, a more comprehensive literature review, in-depth exploration of related works, and a broader range of experiments on segmentation tasks of different targets from different modalities are presented.
We hope that this work will act as catalysts to promote future in-depth research in annotation-efficient medical image segmentation, ultimately benefiting clinical-applicable healthcare practices.

\begin{figure*}[t]
    \centering
	\includegraphics[width=18cm]{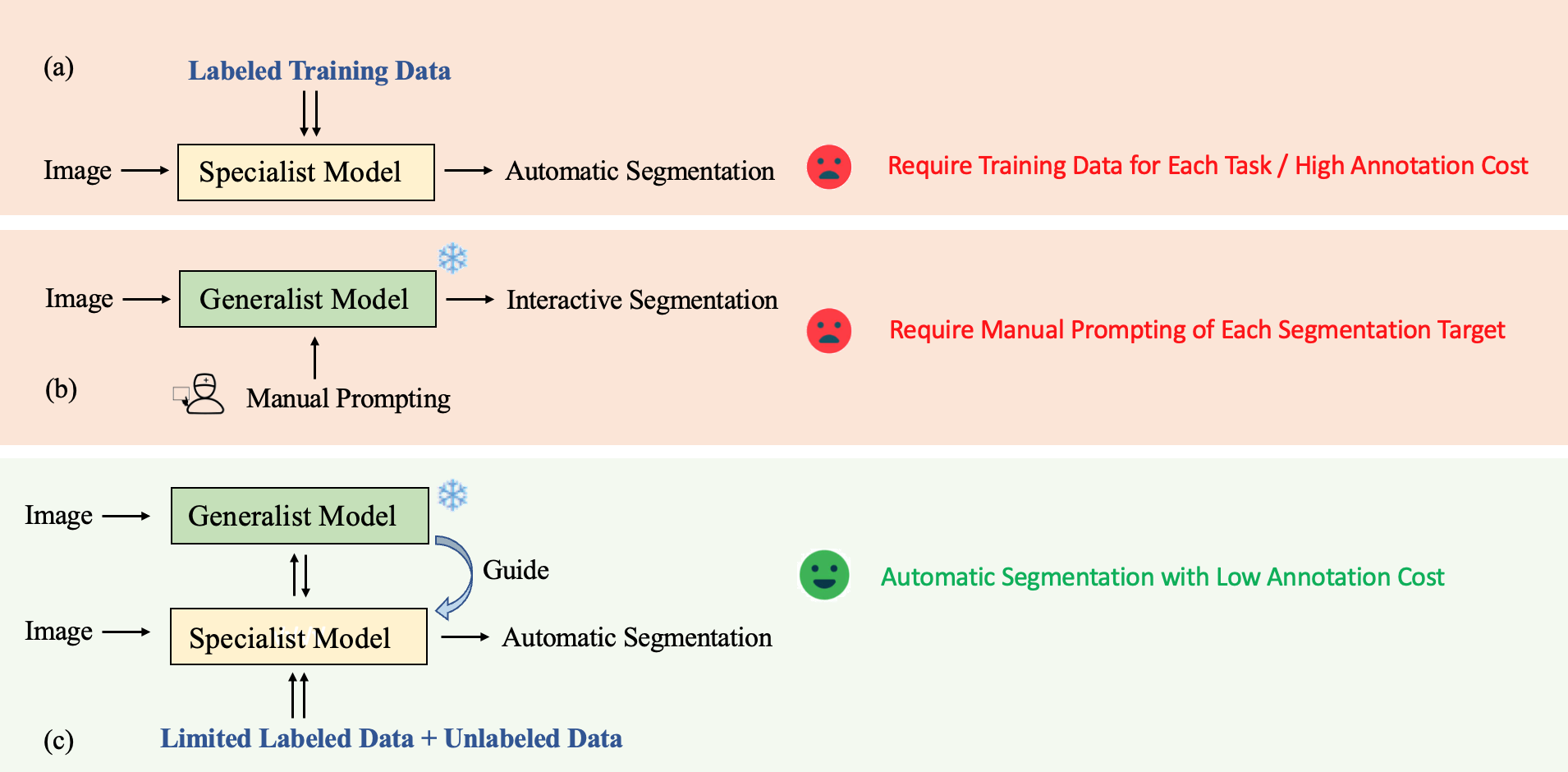}
	\caption{Comparison of existing methods with our proposed method. (a) Specialist model for training-based task-specific automatic segmentation. (b) Generalist model for universal promptable interactive segmentation. (c) SemiSAM+: specialist-generalist collaborative learning for annotation-efficient automatic segmentation.}
    \label{comparison}
\end{figure*}

\section{Related Work}

\subsection{Semi-Supervised Medical Image Segmentation}

To utilize unlabeled data, a direct and intuitive method is assigning pseudo annotations for unlabeled images, and then using the pseudo labeled images in conjunction with labeled images to update the segmentation model \citep{lee2013pseudo}. Existing approaches focus on designing novel generation \citep{seibold2021reference} and selection \citep{huang2022mtl,wang2022ssa} strategies for efficient utilization of pseudo-labeled samples to guide the learning process. 
Other than generating pseudo-labels and updating the segmentation model in an iterative manner, some recent progress of semi-supervised learning aims to enhance model representation learning by designing supervision signals with unsupervised regularization to learn from a substantial amount of unlabeled data. Some research focuses on generating prediction disagreement under different perturbations, like self-ensembling \citep{tarvainen2017mean,yu2019uncertainty}, different resolution scales \citep{luo2022semi}, data recombination \citep{chen2023magicnet}, etc.
Instead of perturbing data for consistency learning, another line of research focuses on building task-level regularization by using adveral training \citep{zhang2017deep,xie2023adversarial}, adding auxiliary task to leverage geometric information \citep{li2020shape,luo2021semi,chen2022semi,zhang2021dual}, or utilizing different learning procedures of different backbone networks \citep{luo2022semi,ma2024semi,song2024sdcl}.
In conclusion, these model-centric advancements aim to exploit more efficient utilization of unlabeled data for improved performance, and the learning procedure is heavily based on exploiting transferable knowledge from labeled images to unlabeled images.
When applying to new segmentation tasks, these semi-supervised methods still need to annotate a small subset of the dataset to obtain acceptable performance.

\subsection{Generalist Foundation Models for Medical Imaging}

Recently, visual foundation models have gained significant attention and have shown impressive performance in medical image analysis \citep{willemink2022toward,moor2023foundation,zhang2024challenges}.
In the field of image segmentation, the Segment Anything Model (SAM) \citep{SAM} introduces the new paradigm of promptable segmentation by interactively segmenting any targets from the image using positional prompts like points and bounding boxes, which demonstrates impressive zero-shot generalization capabilities across various segmentation tasks.

However, several recent studies have revealed SAM’s limited performance in medical image segmentation due to the significant differences between natural images and medical images \citep{SAM-Empirical,SAM4MIS}.
To better adapt SAM for medical imaging, several works focus on adapting SAM to medical images \citep{MedSAM,SAM-Med2D} with improved performance using the same input prompts.
However, these models are still based on the 2D architecture as SAM, which is designed for natural images. To utilize volumetric information of 3D medical images, SAM-Med3D \citep{SAM-Med3D} adopts a SAM-like architecture trained from scratch without utilizing the pre-trained weights of SAM and demonstrates significantly better performance compared with segmenting slice-by-slice, which serves as the default foundation model in our work for 3D medical image segmentation.
Different from universal medical foundation models that combine different modalities and targets for training, some studies also concentrate on developing modality-specific \citep{zhang2025seganypet} or task-specific \citep{gu2025segmentanybone}  foundation models on less universal tasks with better generalization ability.
In conclusion, despite the suboptimal performance and manual efforts required for prompting, these foundation models still opened up new opportunities serving as a reliable generalist models with pre-trained knowledge to guide the learning procedure of downstream tasks.

\begin{figure*}[t]
    \centering
	\includegraphics[width=18cm]{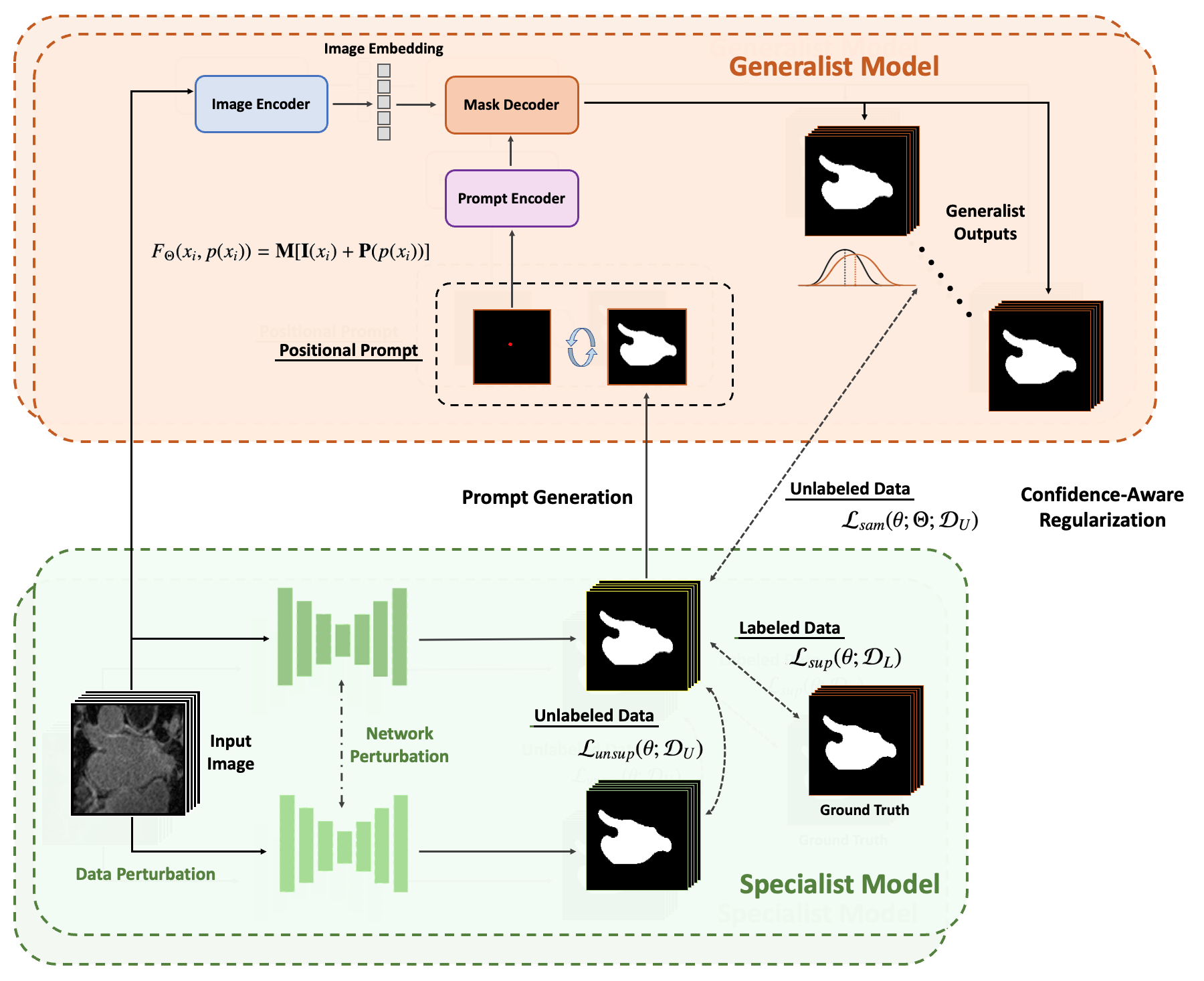}
	\caption{Overview of the proposed foundation model-driven annotation-efficient learning framework \textbf{SemiSAM+}. Specifically, SemiSAM+ consists of a trainable task-specific segmentation model as \textbf{specialist model} for downstream segmentation task and one or multiple promptable foundation models as \textbf{generalist models} pre-trained on large-scale datasets with zero-shot generalization ability. In SemiSAM+, an additional confidence-aware regularization is adapted for effective utilization of generalist model's supervision to avoid possible misguidance when encountering extremely limited annotation.}
	\label{Architecture}
\end{figure*}

\section{Methodology}

\subsection{Preliminaries}

Semi-supervised learning aims to utilize unlabeled data in conjunction with labeled data to train higher-performing segmentation models with limited annotations. 
Given a dataset $\mathcal{D}$ for training, we annotate a small subset with $M$ cases as $\mathcal{D}_{L} = \{x_{l}^{i}, y^{i}\}_{i=1}^{M}$ based on the available budget, while the remaining unlabeled set with $N$ unlabeled cases as $\mathcal{D}_{U} = \{x_{u}^{i}\}_{i=1}^{N}$, where $x_{l}^{i}$ and $x_{u}^{i}$ denote the input images and $y^{i}$ denotes the corresponding ground truth of labeled data. Generally, $\mathcal{D}_{L}$ is a relatively small subset of the entire dataset $\mathcal{D}$, which means $M \ll N$.
For semi-supervised segmentation settings, we aim at building a data-efficient deep learning model to learn with the combination of densely labeled data $\mathcal{D}_{L}$ and unlabeled data $\mathcal{D}_{U}$ for training and to make the performance comparable to an optimal model trained over fully labeled dataset.

Different from generating pseudo-labels and updating the segmentation model in an iterative manner, recent progress in semi-supervised medical image segmentation has been focused on incorporating unlabeled data into the training procedure with unsupervised regularization. 
Generally, the model is optimized based on the combination of supervised segmentation loss $\mathcal{L}_{sup}(\theta;\mathcal{D}_{L})$ and unsupervised regularization loss $\mathcal{L}_{unsup}(\theta;\mathcal{D}_{U})$.
Recent model-centric advancements have been focused on designing novel strategies for calculating unsupervised regularization loss to seek more efficient utilization of unlabeled images.

An overview of the proposed SemiSAM+ framework is illustrated in Fig. \ref{Architecture}. SemiSAM+ consists of a trainable task-specific segmentation model as \textbf{specialist model} for downstream segmentation task (Section \ref{method1}), and one or multiple promptable foundation models as \textbf{generalist models} pre-trained on large-scale datasets with zero-shot generalization ability (Section \ref{method2}). 
Building upon existing specialist SSL frameworks, SemiSAM+ utilizes an additional confidence-aware regularization for effective utilization of generalist model's supervision to avoid possible misguidance and assist in annotation-efficient learning of specialist model (Section \ref{method3}).

\subsection{Specialist: Trainable Task-Specific Segmentation Model} \label{method1}

Building upon the segmentation network for end-to-end fully supervised segmentation using supervised loss, semi-supervised segmentation frameworks mainly focus on generating supervision signal (i.e. unsupervised regularization loss) without requiring annotations based on the unlabeled data.
For the labeled set $D_{L}$, we calculate the supervised segmentation loss $\mathcal{L}_{sup}$ between the main segmentation outputs and the ground truth. While for the unlabeled set $D_{U}$, we calculate the unsupervised regularization loss $\mathcal{L}_{unsup}$ to exploit the information inside the images.
Different semi-supervised segmentation frameworks mainly utilize different strategies to generate unsupervised regularization loss for training with unlabeled data. The total loss for optimization of the specialist model for traditional SSL framework can be organized as follows:
\begin{equation}
\min \limits_{\theta} \mathcal{L}_{sup}(\theta;\mathcal{D}_{L}) + \lambda \mathcal{L}_{unsup}(\theta;\mathcal{D}_{U})
\label{loss}
\end{equation}
Based on different types of regularization, SSL methods can be mainly formulated into several following strategies.

\textbf{Consistency Regularization.}
Consistency regularization is widely applied by enforcing an invariance of predictions of input images under different perturbations, which consists mainly of two components: the main branch with the segmentation network to input original images and generate main segmentation output, and the consistency branch to introduce perturbations to the input images or network conditions to generate additional segmentation output.
These perturbations can be categorized into model perturbations, data perturbations, and simultaneous perturbations of both.
For the unlabeled set, we calculate the unsupervised consistency loss between the main segmentation outputs and addition segmentation outputs, based on the assumption that the segmentation of the same image under different conditions should also be the same.
The definition of unsupervised regularization can be formulated as follows:

\begin{equation}
(\tilde \theta, \tilde x) = \mathcal{T}(\theta,x)
\end{equation}

\begin{equation}
\mathcal{L}_{unsup}(\theta;\mathcal{D}_{U}) =  \mathcal{L}_{con}(S_{\theta}(x_{u}),S_{\tilde \theta}(\tilde x_{u}))
\label{loss_con}
\end{equation}
where $\tilde \theta$ represents the perpetuated model parameters or conditions, and $\tilde x$ represents the perpetuated input image like different augmentations. For some situations with only one kind of perpetuation, $\tilde x = x$ or $\tilde \theta = \theta$.

One of the most representative SSL framework for consistency regularization is the mean teacher (MT) framework \citep{tarvainen2017mean}, where the main branch and the consistency branch are the student model and the teacher model, respectively. The teacher model is an average of student models over different training steps using exponential moving average (EMA) to ensemble weights of the student model at different training stages.
More recent works focus on adding more diverse perturbations like different network backbones (e.g. Transformer \citep{luo2021crossteaching} and Mamba \citep{ma2024semi}) and foreground-background copy-paste between different images \citep{song2024sdcl}.

\textbf{Region-Selective Regularization.}
Instead of directly utilizing the whole image for consistency regularization, several work focus on utilizing the estimated confidence (i.e. uncertainty) of model-generated predictions to guide the regularization.
For unsupervised regularization, similar to Equation \ref{loss_con}, this strategy re-design the consistency loss as follows:
\begin{equation}
\mathcal{L}_{rs-con}(f_{1},f_{2},M) = \frac{ \sum ( M \| f_{1} - f_{2} \|^{2}  )  }{ \sum M }
\end{equation}
where $M$ is region-selective mask as a binary ambiguity indicator, $f_{1}$ and $f_{2}$ are the output predictions for consistency regularization like $S_{\theta}(x_{u})$ and $S_{\tilde \theta}(\tilde x_{u})$. With region-selective regularization, the model can directly focus on learning the most informative clues from the unlabeled images based on their consensus on these regions. 

One of the most representative SSL framework for this strategy is the uncertainty-aware mean teacher (UA-MT) framework \citep{yu2019uncertainty} by filtering out unreliable regions and preserving only reliable regions with low uncertainty when calculating the consistency loss.
With the guidance of uncertainty, the model gradually learns from the meaningful and reliable information from the unlabeled data.
Based on estimated uncertainty map $U$, the region-selective mask $M$ is generated based on a pre-defined uncertainty threshold $u_{th}$.
\begin{equation}
M =  \mathbb I(U<u_{th}) 
\end{equation}
Besides, another framework name ambiguity-consensus mean-teacher (AC-MT) \citep{xu2023ambiguity} further extend the generation strategy of region-selective mask to more perspectives including high softmax entropy and label noise self-identification. Uncertainty-based region selection is also adapted in other SSL frameworks with other types of regularization. \citep{zhang2023uncertainty,adiga2024anatomically}.

\textbf{Auxiliary Task.}
Instead of data-level perturbations for unsupervised regularization, another line of researches focus on exploiting various levels of information from different tasks with different focuses by incorporating auxiliary task that capture different aspects of the data.
One of the most representative framework for auxiliary task is the Dual Task Consistency (DTC) framework \citep{luo2021semi}, where the model consists of two different tasks.
Beyond the traditional pixel/voxel-level classification task for image segmentation, the model adopts another branch to predict geometry-aware level set representations of the segmentation target.
The dual-task consistency strategy encourages the model to produce consistent predictions after the transformation of different tasks.
The definition of unsupervised regularization can be formulated as follows:
\begin{equation}
\mathcal{L}_{unsup}(\theta;\mathcal{D}_{U}) =  \mathcal{L}_{con}(S_{\theta}(x_{u}),\mathcal{T} (R_{\tilde \theta}(x_{u})))
\end{equation}
where $S$ and $\theta$ represent the model and parameter of classification task, while $R$ and $\tilde \theta$ represent the model and parameter of additional level-set regression task. $\mathcal{T}$ represents the transformation of different tasks into the same space for regularization.

\textbf{Adversarial Training.}
Adversarial training is a technique used to improve the robustness and generalization of models by exposing them to adversarial examples, which are inputs that have been perturbed in a way intended to fool the model. 
One of the most representative SSL frameworks for adversarial training is the Deep Adversarial Network (DAN) framework \citep{zhang2017deep}.
The DAN framework consists of two main components: a segmentation network (SN) and an evaluation network (EN). The segmentation network is responsible for generating segmentation masks from input images, while the evaluation network assesses the quality of these segmentations. The adversarial training process involves an iterative game between these two networks.
The segmentation network is trained to produce segmentations that are indistinguishable from those of annotated images, even when the input comes from unannotated images.
The evaluation network is trained to differentiate between segmentations produced from annotated and unannotated images, assigning higher scores to those that appear to be of higher quality.

During model training, EN takes the segmentation probability maps and the corresponding image as input to determine a score indicating the quality of the segmentation, and is encouraged to give high scores (1) for labeled data and low scores (0) for unlabeled data.
The training of EN aims to minimize the following function:
\begin{equation}
\mathcal{L}_{EN} = \mathcal{L}_{bce}(E_{\tilde \theta}(S_{\theta}(x_{l}),x_{l}),1) + \mathcal{L}_{bce}(E_{\tilde \theta}(S_{\theta}(x_{u}),x_{u}),0)
\end{equation}
where $S_{\theta}$ and $E_{\tilde \theta}$ represent the segmentation network and an evaluation network.
For the training of segmentation network, with respect to Equation \ref{loss}, the definition of unsupervised regularization can be formulated as follows:
\begin{equation}
\mathcal{L}_{unsup}(\theta;\mathcal{D}_{U}) =  - \mathcal{L}_{bce}(E_{\tilde \theta}(S_{\theta}(x_{u}),x_{u}),0)
\end{equation}

\textbf{Summary.} 
As stated above, these model-centric SSL advancements aim to design different strategies for unsupervised regularization to exploit unlabeled data. In this work, we investigate these four most representative SSL frameworks of different strategies as specialist models for SemiSAM+ in the following experiments.

\subsection{Generalist: SAM-like Segmentation Foundation Model}
 \label{method2}

As the first promptable foundation model for image segmentation, SAM \citep{SAM} represents a new paradigm of promptable segmentation for universal segmentation.
Specifically, SAM adopts an image encoder to extract image embeddings, a prompt encoder to integrate user interactions via different prompt modes, and a mask decoder to predict segmentation masks by fusing image embeddings and prompt embeddings. 
Building upon the promptable segmentation foundation model, we can use one model to segment anything from a given image based on the positional prompt as shown in Fig. \ref{SAM}.
The overall workflow of SAM for promptable segmentation can be defined as follows:
\begin{equation}
F_{\Theta}(x_{i},p(x_{i}) ) = \mathbf{M}[ \mathbf{I}(x_{i}) + \mathbf{P}(p(x_{i})) ]
\label{eq_sam}
\end{equation}
where $\mathbf{I}$, $\mathbf{P}$, and $\mathbf{M}$ represent the image encoder, the prompt encoder, and the mask decoder, respectively. $p(x_{i})$ represents the positional prompting of the input image $x_{i}$.

Despite the original SAM, several works have also focused on adapting the promptable segmentation paradigm into universal medical image segmentation tasks.
For example, MedSAM \citep{MedSAM} adapts the original SAM to medical image segmentation by curating a diverse and comprehensive 2D medical dataset.
However, the structure still follows the original SAM and only processes 3D images as a series of 2D slices instead of the volumes.
The inherent 2D architecture often leads to sub-optimal results due to the lack of inter-slice spatial context in 3D medical images, which is extremely important for identification of some objects so as to ensure accurate segmentation \citep{zhang2022bridging}.
To address this issue, many studies have undertaken specific modifications and enhancements for 3D promptable segmentation foundation models represented by SAM-Med3D \citep{SAM-Med3D}.
The SAM-Med3D model with a learnable 3D SAM-like architecture is trained from scratch on a large-scale 3D medical dataset and supports point and mask prompts.
Besides, several other modality-specific (e.g. SegAnyPET \citep{zhang2025seganypet}) and task-specific (e.g. SegmentAnyBone \citep{gu2025segmentanybone}) generalist models focus on promptable segmentation of less universal tasks with better generalization ability.
In this work, we use SAM-Med3D as the default generalist model in the following experiments. While SegAnyPET is used for PET segmentation task to evaluate the feasibility of SemiSAM+ for other and multiple generalist models.

\begin{table}[]
	\centering
    \normalsize
    \setlength\tabcolsep{6pt}
	\renewcommand\arraystretch{1.3}
     \caption{Information summary of datasets used in our work.} \label{Table_Dataset}
	\begin{tabular}{c|ccc}
		\hline
Information & Dataset &  Modality  & Target  \\ \hline
\multirow{2}{*}{Public}  & LA &  GE-MRI & Left Atrium  \\ 
 & BraTS19 &  FLAIR-MRI & Brain Tumor  \\  \hline
In-house   & PETS & FDG-PET & Whole Heart \\ \hline 
\end{tabular}
\end{table}

\begin{table*}[t]
	\caption{Comparison of segmentation performance with generalist models and specialist models on different segmentation datasets. Specifically, N denotes the count of slices containing the target object.} \label{Table_compare}
	\centering
    \small
	\renewcommand\arraystretch{1.4}
	\begin{tabular}{c|c|c|c|cc}
		\hline 	\hline
		\multirow{2}{*}{Model Type} & \multirow{2}{*}{Method}  & \multirow{2}{*}{Manual Annotation}  & \multirow{2}{*}{Manual Prompt} &  \multicolumn{2}{c}{Dice Performance $\uparrow$[\%]}  \\
        \cline{5-6}		& &&& Left Atrium & Brain Tumor  \\ \hline
          \multirow{8}{*}{Specialist Model} &  \multirow{4}{*}{FS Baseline}   & 1 labeled case    & \multirow{4}{*}{-}   & 37.97  & 42.82      \\
               && 2 labeled cases  && 59.99  & 53.70   \\
               && 3 labeled cases  && 62.33  & 56.06   \\
               && 5 labeled cases  && 63.39  & 62.35   \\
           \cline{2-6} &  \multirow{4}{*}{SSL MT}   & 1 labeled case    & \multirow{4}{*}{-}    & 40.64  & 52.56      \\
               && 2 labeled cases   && 61.84  & 66.16   \\
               && 3 labeled cases   && 64.35  & 68.11   \\
               && 5 labeled cases   && 64.82  & 70.48   \\ \hline
        \multirow{8}{*}{Generalist Model} & \multirow{4}{*}{SAM}         & \multirow{4}{*}{-}                         & N points             & 41.98   & 32.85    \\
                     & &  & 2N points           & 46.11    & 39.41  \\
                     & &  & 5N points           & 58.90    & 60.35      \\
                     & &  & 10N points           & 71.38    & 79.86       \\ 
		   \cline{2-6} & \multirow{4}{*}{SAM-Med3D }   & \multirow{4}{*}{-}              & 1 points             & 56.51    &  53.91         \\
               &  & & 2 points             & 63.96    & 62.05     \\
               &  &  & 5 points             & 70.28    &   68.23        \\
               &  & & 10 points            & 73.78    & 74.99          \\  \hline
              &  \multirow{4}{*}{SSL MT / SAM-Med3D}   & 1 labeled case        & \multirow{4}{*}{-}                     &  48.91 & 68.53 \\
         Specialist-Generalist Collaboration & & 2 labeled cases          &  & 73.09 & 72.59     \\  
          (SemiSAM+) & & 3 labeled cases            & & 75.66 & 73.92 \\ 
                       & & 5 labeled cases            & & 77.02 & 76.00  \\  \hline
           Upperbound & FS Upperbound   & All labeled cases  & -                    & 88.18   & 87.29                 \\  \hline \hline
	\end{tabular}
\end{table*}

\begin{table*}[t]
	\caption{Ablation study of the proposed SemiSAM+ framework on the LA dataset with different prompting strategy. (m) directly utilizing specialist output as mask prompt. (p) generating point prompts based on the specialist output. (ur) uncertainty-rectified confidence-aware regularization.} \label{Table_ablation}
	\centering
    \small
    \setlength\tabcolsep{10pt}
	\renewcommand\arraystretch{1.3}
	\begin{tabular}{c|c|cccc}
		\hline 	\hline
        Method	& $L/U$ & Dice$\uparrow$[\%] & Jaccard$\uparrow$[\%] & 95HD$\downarrow$[voxel] & ASD$\downarrow$[voxel]  \\ \hline 
          FS Baseline  & 1 / 0  & 37.97 & 25.83 & 39.50 & 16.46 \\
          SSL Baseline & 1 / 74 & 40.64 & 28.42 & 38.92 & 15.85  \\ \hline
          SemiSAM (m) & 1 / 74 & 48.78 & 35.41 & 38.22 & 14.36 \\ 
          SemiSAM (p) & 1 / 74 & 47.94 & 34.74 & 38.49 & 14.79 \\ 
          SemiSAM+ (ur) & 1 / 74 & 48.91 & 35.46 & 38.49 & 14.61\\ \hline \hline
          FS Baseline   & 2 / 0  & 59.99 & 45.02 & 38.44 & 14.11     \\ 
          SSL Baseline & 2 / 73  & 61.84 & 46.69 & 38.65 & 13.86 \\ \hline
          SemiSAM (m) & 2 / 73  & 72.70 & 58.73 & 31.61 & 10.29 \\ 
          SemiSAM (p) & 2 / 73  & 72.96 & 58.85 & 35.83 & 11.32 \\
          SemiSAM+ (ur) & 2 / 73  & 73.09 & 59.01 & 34.05 & 10.83 \\ \hline \hline
          FS Baseline   & 3 / 0  & 62.33 & 47.19 & 36.59 & 12.16       \\ 
          SSL Baseline & 3 / 72 & 64.35 & 49.47 & 31.40 & 10.93 \\ \hline
          SemiSAM (m) & 3 / 72 & 75.43 & 62.69 & 23.30 & 7.51 \\ 
          SemiSAM (p) & 3 / 72 & 74.64 & 61.72 & 26.40 & 8.50 \\ 
          SemiSAM+ (ur) & 3 / 72 & 75.66 & 62.36 & 22.25 & 6.59 \\ \hline \hline
          FS Baseline   & 5 / 0 & 63.39 & 48.91 & 36.10 & 11.91      \\ 
          SSL Baseline & 5 / 70 & 64.82 & 50.07 & 34.26 & 11.47 \\ \hline
          SemiSAM (m) & 5 / 70 & 76.09 & 63.14 & 22.17 & 6.95 \\ 
          SemiSAM (p) & 5 / 70 & 75.14 & 62.03 & 24.45 & 7.96 \\
          SemiSAM+ (ur) & 5 / 70 & 77.02 & 64.33 & 23.30 & 7.48 \\ \hline \hline
          FS Upperbound  & 75 / 0  & 88.18 & 79.23 & 11.34 & 2.90   \\ \hline \hline
	\end{tabular}
\end{table*}

\subsection{SemiSAM+: Foundation Model-Driven Annotation-Efficient Medical Image Segmentation}
\label{method3}

Despite existing specialist semi-supervised models can utilize unlabeled data to improve the model performance, they still struggle to achieve satisfactory outcomes when only extremely limited annotations are available.
One possible reason is that the model cannot learn sufficient discriminative information with extremely limited supervision. As a result, the predictions are full of randomness and the improvement of exploiting unlabeled data is limited.
To issue this challenge, we aim to leverage the pre-trained knowledge base of these generalist foundation models to guide the learning procedure of specialist model when annotations are scarce.

As shown in Fig. \ref{Architecture}, the training procedure of SemiSAM+ consists of a specialist-generalist collaborative learning where the trainable specialist model delivers positional prompts to interact with the frozen generalist model to acquire pseudo segmentation, and then the generalist model output provides the specialist model with informative and efficient supervision which benefits the automatic segmentation and prompt generation process in turn.
For universal promptable segmentation settings, the segmentation output primarily relies on the input positional prompts. Different from the interactive semi-automatic segmentation manner of manual prompting, in SemiSAM+, the segmentation output of the trainable specialist model $S_{\theta}(x_{i})$ can be utilized to provide information for generating positional prompt $p(x_{i})$ for generalist models. 
Despite the unsatisfying segmentation when only limited annotations are available, the segmentation output is sufficient for coarse localization and positional prompting of the segmentation target.
Specifically, we introduce different prompting strategies including directly utilizing the coarse segmentation as mask prompt and generating point prompts based on the coarse segmentation.
The acquisition of positional prompt can be defined as follows:
\begin{equation}
p_{n}(x_{i}) = Prompting_{n} ( S_{\theta}(x_{i}) )
\end{equation}
After that, the generalist model can then generate segmentation outputs based on the input image $x_{i}$ and positional prompt $p_{n}(x_{i})$ as Equation \ref{eq_sam}.
Based on the positional prompting, the generalist model's output can provide the specialist model with informative and efficient supervision, which in turn benefits the automatic segmentation and prompt generation.
To utilize the knowledge from generalist model and assist in the learning process of specialist model, we add an additional supervision branch upon the classic SSL framework.

As observed in our previous work \citep{zhang2024semisam}, the improvement of simply enforcing the consistency of outputs between the specialist model and the generalist model is limited when a relatively large amount of labeled data is available.
In SemiSAM+, we utilize a confidence-aware regularization for effective utilization of generalist models' knowledge to assist in annotation-efficient learning of specialist model.
Along with the prediction, we aim to estimate the uncertainty of the output of generalist model to assist in the regularization.
Building upon the foundation of promptable segmentation, we leverage the advantage of the prompting mechanism and approximate the segmentation uncertainty with different prompts. The aleatoric uncertainty of generalist model is formulated as follows:
\begin{equation}
U_{x} = \mathit{D} [ F_{\Theta}(x,p_{1}(x) ) , ... , F_{\Theta}(x,p_{n}(x))]
\end{equation}
where $\mathit{D}$ computes the statistical discrepancy of predictions using different input prompts to estimate the pixel/voxel- level uncertainty $U_{x}$.
Specifically, the estimation procedure does not require specific designs of the specialist model or generalist model with strong flexibility.
Then the confidence-aware regularization can be organized as follows:
\begin{equation}
\mathcal{L}_{sam}(\theta;\Theta;\mathcal{D}_{U}) =  \mathcal{L}_{rs-con}(S_{\theta}(x_{u}),F_{\Theta}(x_{u},p(x_{u})),U_{x})
\label{loss_sam}
\end{equation}
Finally, the overall training loss for SemiSAM+ can be formulated as follow:
\begin{equation}
\min \limits_{\theta} \mathcal{L}_{sup}(\theta;\mathcal{D}_{L}) + \lambda \mathcal{L}_{unsup}(\theta;\mathcal{D}_{U}) + \beta \mathcal{L}_{sam}(\theta;\Theta;\mathcal{D}_{U})
\label{final_loss}
\end{equation}
where $\lambda$ and $\beta$ are weighting coefficients for different components.

\subsection{Extend to Multiple Generalist Models}

Due to the different learning aspects of different generalist models, we can extend the learning procedure of SemiSAM+ with multiple generalists to more reliable guidance.
Given n generalist models with parameters of $\Theta_{1},...,\Theta_{n}$, the overall training loss can be extended as follows:
\begin{equation}
\min \limits_{\theta} \mathcal{L}_{sup}(\theta;\mathcal{D}_{L}) + \lambda \mathcal{L}_{unsup}(\theta;\mathcal{D}_{U}) + \beta \sum_{i=1}^{n} 
\mathcal{L}_{sam}(\theta;\Theta_{i};\mathcal{D}_{U})
\label{final_loss}
\end{equation}

\begin{table*}[t]
	\caption{Evaluation of flexibility of adapting SemiSAM+ upon different specialist models on left atrium segmentation and brain tumor segmentation tasks under different annotation scenarios. All the models use the same 3D U-Net as the backbone.} \label{Table_specialist}
	\centering
    \scriptsize
	\renewcommand\arraystretch{1.3}
	\begin{tabular}{c|c|cccc|c|cccc}
		\hline 	\hline
		 \multirow{2}{*}{Method}  &   \multicolumn{5}{c|}{Left Atrium Segmentation} &   \multicolumn{5}{c}{Brain Tumor Segmentation} \\
        \cline{2-11}		& $L/U$ & Dice$\uparrow$[\%] & Jaccard$\uparrow$[\%] & 95HD$\downarrow$[voxel] & ASD$\downarrow$[voxel] & $L/U$ & Dice$\uparrow$[\%] & Jaccard$\uparrow$[\%] & 95HD$\downarrow$[voxel] & ASD$\downarrow$[voxel] \\ \hline 
          FS Baseline   & 1 / 0  & 37.97 & 25.83 & 39.50 & 16.46 & 1 / 0 & 42.82 & 29.01 & 63.55 & 30.43       \\ \hline
          1) MT & 1 / 74 & 40.64 & 28.42 & 38.92 & 15.85 & 1 / 249 & 52.56 & 37.13 & 58.80 & 25.41 \\
          SemiSAM+ & 1 / 74 & 48.91 & 35.46 & 38.49 & 14.61 & 1 / 249 & 68.53 & 56.22 & 38.81 & 16.66 \\ \hline
          2) UA-MT & 1 / 74 & 41.73 & 29.46 & 39.74 & 15.81 & 1 / 249 & 65.38 & 51.61 & 46.50 & 19.85 \\
          SemiSAM+ & 1 / 74 & 50.10 & 36.56 & 37.02 & 14.27 & 1 / 249 & 71.75 & 58.71 & 34.24 & 12.91 \\ \hline
          3) DAN & 1 / 74 & 44.61 & 31.08 & 38.03 & 14.68 & 1 / 249 & 62.68 & 49.45 & 46.79 & 20.61 \\
          SemiSAM+ & 1 / 74 & 62.26 & 46.82 & 31.88 & 11.94 & 1 / 249 & 70.32 & 57.93 & 34.18 & 13.78 \\ \hline
          4) DTC & 1 / 74 & 41.62 & 28.76 & 38.08 & 15.02 & 1 / 249 & 53.48 & 39.21 & 33.20 & 12.07 \\
          SemiSAM+ & 1 / 74 & 53.52 & 39.84 & 35.19 & 13.25 & 1 / 74 & 67.29 & 55.19 & 30.82 & 11.90 \\ \hline \hline
          FS Baseline   & 2 / 0  & 59.99 & 45.02 & 38.44 & 14.11 & 2 / 0 & 53.70 & 38.98 & 77.37 & 33.57       \\ \hline
          1) MT & 2 / 73  & 61.84 & 46.69 & 38.65 & 13.86 & 2 / 248 & 66.16 & 52.21 & 45.73 & 19.07 \\
          SemiSAM+ & 2 / 73 & 73.09 & 59.01 & 34.05 & 10.83 & 2 / 248 & 72.59 & 60.04 & 37.57 & 14.32 \\ \hline
          2) UA-MT & 2 / 73 & 62.37 & 47.46 & 39.12 & 13.92 & 2 / 248 & 67.07 & 52.62 & 40.59 & 15.57 \\
          SemiSAM+ & 2 / 73 & 74.06 & 60.26 & 30.75 & 9.81 & 2 / 248 & 72.26 & 60.04 & 33.14 & 13.43 \\ \hline
          3) DAN & 2 / 73 & 59.27 & 43.33 & 41.71 & 15.18 & 2 / 248 & 66.45 & 52.73 & 37.53 & 14.04 \\
          SemiSAM+ & 2 / 73 & 72.28 & 58.46 & 35.29 & 11.42 & 2 / 248 & 72.54 & 60.34 & 22.47 & 8.48 \\ \hline
          4) DTC & 2 / 73 & 60.14 & 45.18 & 40.22 & 14.05 & 2 / 248 & 59.16 & 43.97 & 45.91 & 20.11 \\
          SemiSAM+ & 2 / 73 & 73.76 & 60.34 & 29.56 & 9.72 & 2 / 248 & 70.96 & 58.48 & 33.26 & 13.93 \\ \hline \hline
          FS Baseline   & 3 / 0  & 62.33 & 47.19 & 36.59 & 12.16 & 3 / 0 & 56.06 & 42.81 & 47.67 & 17.52      \\ \hline
          1) MT & 3 / 72 & 64.35 & 49.47 & 31.40 & 10.93 & 3 / 247 & 68.11 & 54.45 & 36.57 & 12.85 \\
          SemiSAM+ & 3 / 72 & 75.66 & 62.36 & 22.25 & 6.59 & 3 / 247 & 73.92 & 62.51 & 19.77 & 6.13  \\ \hline
          2) UA-MT & 3 / 72 & 65.18 & 50.68 & 31.37 & 10.73 & 3 / 247 & 67.28 & 53.45 & 32.14 & 9.77 \\
          SemiSAM+ & 3 / 72 & 75.97 & 63.04 & 24.78 & 7.82 & 3 / 247 & 73.86 & 62.38 & 21.46 & 7.18 \\ \hline
          3) DAN & 3 / 72 & 63.70 & 48.85 & 36.26 & 12.40 & 3 / 247 & 68.37 & 54.48 & 38.52 & 15.07 \\
          SemiSAM+ & 3 / 72 & 73.67 & 60.72 & 21.83 & 7.72 & 3 / 247 & 74.52 & 62.68 & 21.22 & 6.56 \\ \hline
          4) DTC & 3 / 72 & 64.44 & 50.86 & 24.38 & 7.85 & 3 / 247 & 61.96 & 48.07 & 28.95 & 8.43 \\
          SemiSAM+ & 3 / 72 & 73.16 & 60.38 & 21.22 & 6.76 & 3 / 247 & 75.30 & 63.76 & 19.83 & 5.27 \\ \hline \hline
          FS Baseline   & 5 / 0 & 63.39 & 48.91 & 36.10 & 11.91 & 5 / 0 & 62.35 & 50.16 & 36.68 & 11.83    \\ \hline
          1) MT & 5 / 70 & 64.82 & 50.07 & 34.26 & 11.47 & 5 / 245 & 70.48 & 57.91 & 35.79 & 11.83 \\
          SemiSAM+ & 5 / 70 & 77.02 & 64.33 & 23.30 & 7.48 & 5 / 245 & 76.00 & 63.72 & 25.82 & 8.89\\ \hline
          2) UA-MT & 5 / 70 & 65.51 & 50.77 & 35.51 & 11.46 & 5 / 245 & 68.51 & 56.39 & 37.90 & 13.40 \\
          SemiSAM+ & 5 / 70 & 74.92 & 61.65 & 24.06 & 7.82 & 5 / 245 & 74.05 & 62.61 & 22.61 & 6.55 \\ \hline
          3) DAN & 5 / 70 & 65.09 & 50.22 & 33.99 & 11.18 & 5 / 245 & 69.32 & 56.46 & 34.73 & 12.19 \\
          SemiSAM+ & 5 / 70 & 75.25 & 61.93 & 24.24 & 7.55 & 5 / 245 & 74.97 & 62.56 & 27.88 & 10.24 \\ \hline
          4) DTC & 5 / 70 & 68.04 & 53.90 & 24.76 & 7.87 & 5 / 245 & 65.03 & 51.73 & 30.58 & 8.87 \\
          SemiSAM+ & 5 / 70 & 79.52 & 67.33 & 19.03 & 5.54 & 5 / 245 & 76.53 & 64.82 & 20.44 & 6.52 \\ \hline \hline
          FS Upperbound & 75 / 0  & 88.18 & 79.23 & 11.34 & 2.90 & 250 / 0 & 87.29 & 78.55 & 7.77 & 1.73  \\ \hline \hline
	\end{tabular}
\end{table*}

\section{Experiments}

\subsection{Dataset and Implementation Details}

To evaluate the performance of the proposed SemiSAM+ framework, we conduct experiments on two public datasets and one in-house clinical dataset. An information summary of the datasets used in our work is shown in Table. \ref{Table_Dataset}. All these datasets are 3D segmentation tasks.

\textbf{Public benchmark datasets.}
The first dataset is the Left Atrium Segmentation Challenge Dataset \citep{xiong2021global}.
The dataset contains 100 3D gadolinium-enhanced MR imaging scans (GE-MRIs) and corresponding LA segmentation masks for training and validation. These scans have an isotropic resolution of $0.625 \times 0.625 \times 0.625 mm^{3}$. 
We divide the 100 scans into 75 scans for training, 5 scans for validation, and 20 scans for testing. 
The second dataset is the BraTS 2019 dataset \citep{hdtd-5j88-20} for brain tumor segmentation.
We investigate semi-supervised segmentation of the whole tumors from FLAIR MRI images following the design in \citep{luo2022semi} since this modality can well characterize the malignant tumors \citep{zeineldin2020deepseg}. We divide the dataset into 250 scans for training, 25 scans for validation, and 60 scans for testing.

\textbf{In-house clinical dataset.}
To further validate the experiments in real clinical scenarios, we conduct additional experiments on one in-house clinical dataset.
The dataset contains 100 FDG-PET images with expert-examined annotations of the heart for quantitative evaluation of the cardiac metabolism.
The PET images were reconstructed with 4.07 $\times$ 4.07 $\times$ 3 $mm^{3}$ voxels.
We divide the 100 scans into 40 scans for training, 10 scans for validation, and 50 scans for testing.

\begin{table*}[t]
	\caption{Comparison of Dice performance of SemiSAM+ with state-of-the-art semi-supervised and SAM-based fine-tuning methods on the LA dataset.} \label{Table_sota}
	\centering
    \small
	\renewcommand\arraystretch{1.4}
	\begin{tabular}{c|c|cccc}
		\hline 	\hline
        Model Type & Method	& 1 labeled & 2 labeled & 3 labeled & 5 labeled \\ \hline 
        FS Baseline  & 3D U-Net \citep{cciccek20163d} & 37.97 & 59.99 & 62.33 & 63.39 \\ \hline 
        \multirow{5}{*}{SAM-based} & AutoSAM \citep{autosam} & 57.99 & 67.21 & 72.12 & 72.53\\
        & SAMed \citep{SAMed} & 64.16 & 74.71 & 75.04 & 77.38 \\
        & MA-SAM \citep{chen2024ma} & 36.67 & 40.88 & 58.76 & 74.52 \\
        & H-SAM \citep{cheng2024unleashing} & 67.73 & 75.86 & 78.36 & 81.82 \\  
        & Auto-SAM-Med3D & 61.07 & 68.95 & 72.92 & 75.33 \\ \hline
        \multirow{4}{*}{SSL methods}
        & UGMCL \citep{zhang2023uncertainty} & 44.08 & 66.04 & 69.05 & 70.22 \\
        & ACMT \citep{xu2023ambiguity} & 44.17 & 71.74 & 73.45 & 74.84 \\ 
        \cline{2-6} & MT \citep{tarvainen2017mean} & 40.64 & 61.84 & 64.35 & 64.82 \\
        & SDCL* \citep{song2024sdcl} & 59.75 & 80.79 & 87.67 & 88.04 \\ \hline
        \multirow{2}{*}{SemiSAM+ } & MT / SAM-Med3D & 48.91 & 73.09 & 75.66 & 77.02  \\ 
         & SDCL* / SAM-Med3D & 76.95 & 86.37 & 88.12 & 90.14 \\ \hline \hline
	\end{tabular}
    \\
    $*$ denotes the method is implemented with multiple network backbones different from other SSL implementations.
\end{table*}

\begin{figure*}[h]
	\includegraphics[width=18cm]{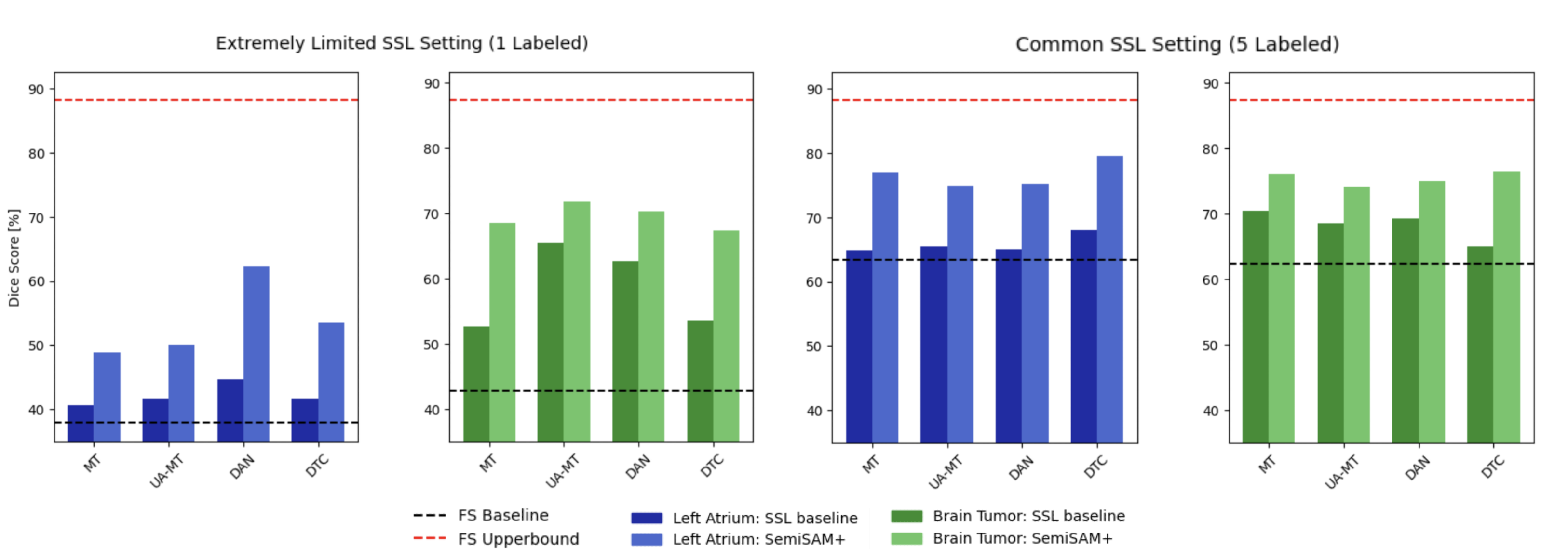}
	\caption{Comparison of Dice performance of adapting SemiSAM+ to different SSL specialist models for left atrium segmentation and brain tumor segmentation under different SSL settings.}
    \label{plot}
\end{figure*}

\begin{table*}[h]
	\caption{Comparison of Dice performance of adapting SemiSAM+ upon different single and multiple generalist models on the in-house PETS dataset.} \label{Table_pet}
	\centering
    \small
	\renewcommand\arraystretch{1.4}
	\begin{tabular}{c|c|cccc}
		\hline 	\hline
        Model Type & Method	&  1 labeled & 2 labeled & 3 labeled & 5 labeled \\ \hline 
        FS Baseline  & 3D U-Net \citep{cciccek20163d} & 27.66 & 28.51 & 30.47 & 31.51 \\ \hline 
        SSL Baseline & MT \citep{tarvainen2017mean} & 35.68 & 40.54 & 41.91 & 59.26 \\ \hline
        SemiSAM+ & MT / SAM-Med3D &  34.24 & 59.76 & 62.65 & 67.34 \\ 
        Single Specialist & MT / SegAnyPET & 45.33 & 63.07 & 64.71 & 67.74\\ \hline
        SemiSAM+ & \multirow{2}{*}{MT / SAM-Med3D+SegAnyPET} & \multirow{2}{*}{39.54} & \multirow{2}{*}{60.50} & \multirow{2}{*}{67.61} & \multirow{2}{*}{72.88} \\
        Multiple Specialists && \\ \hline \hline
	\end{tabular}
\end{table*}

\begin{figure}
	\includegraphics[width=9cm]{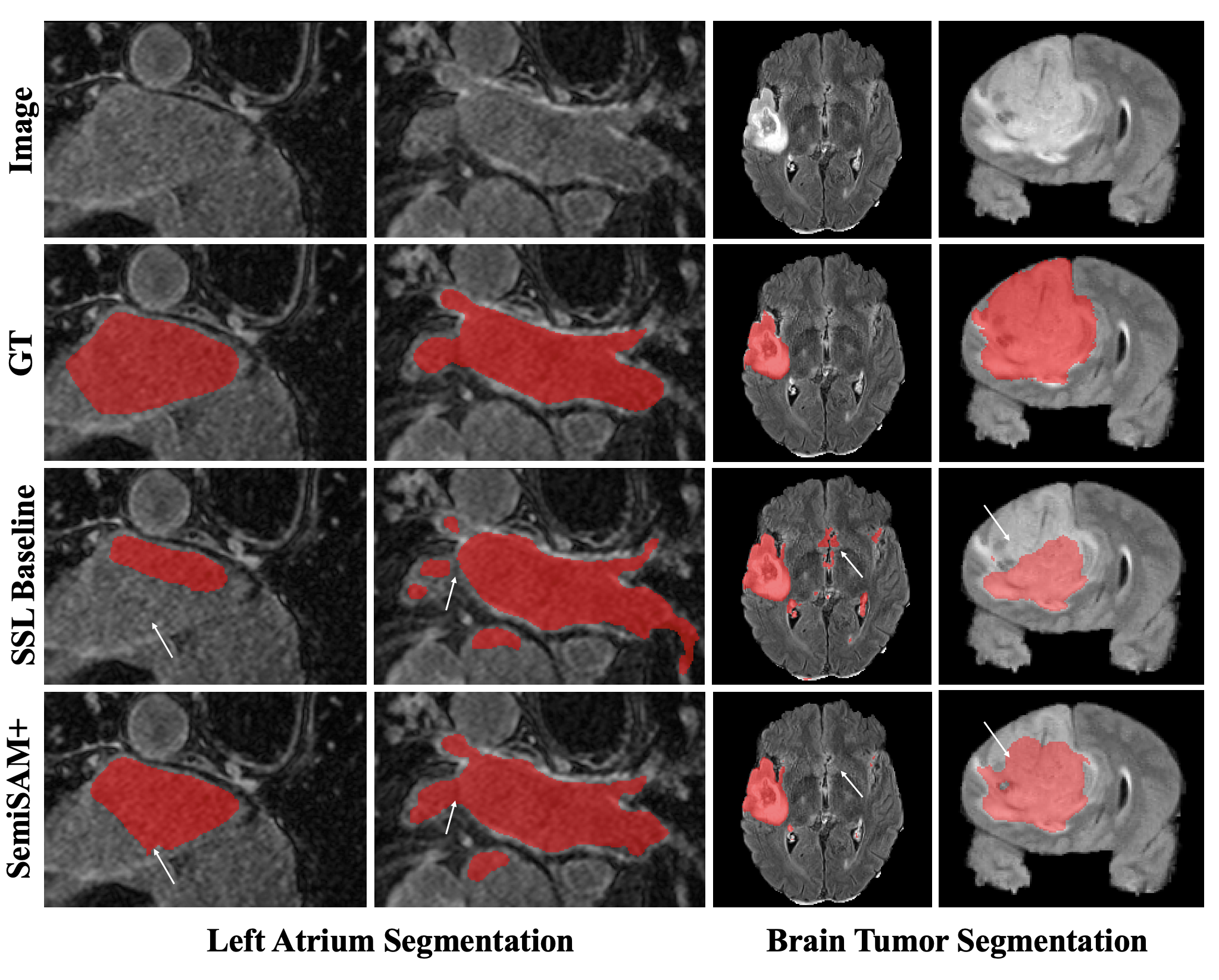}
	\caption{Visual comparison of segmentation results of SSL baseline (MT) and SemiSAM+ (MT / SAM-Med3D) on left atrium segmentation and brain tumor segmentation.}
    \label{vis_labt}
\end{figure}

\textbf{Inplementation details.}
All of our experiments are implemented in Python with PyTorch, using an NVIDIA A100 GPU. The backbone segmentation network for the specialist model is 3D U-NET \citep{cciccek20163d}. 
We use the SGD optimizer with an initial learning rate of 0.01, a weight decay of 1e-4 and a momentum of 0.9 to update the network parameters with the maximum iteration number set to 30000.
The poly learning rate strategy is used to adjust the learning rate, where the initial learning rate was multiplied by $(1.0 - \frac{t}{t_{max}})^{0.9}$.
Due to the extremely limited annotation scenario, the batch size is set to 2 for all the compared methods, where half of the batch are labeled data and the other half are unlabeled images. The model takes randomly cropped patches as input.
The patch size of the specialist model is set to $128 \times 128 \times 128$ sub-volumes to fit in the input shape of generalist model in our experiment. During the training stage, random cropping, flipping and rotation are used to enlarge the training set and avoid over-fitting. In the inference stage, the final segmentation results are obtained using a sliding-window strategy. 
We use four commonly used metrics in segmentation tasks for evaluation, including the Dice similarity coefficient (Dice), the Jaccard index (Jaccard), 95\% Hausdorff Distance (95HD), and the Average Surface Distance (ASD). Lower 95HD and ASD indicate better segmentation performance, while larger Dice and Jaccard indicate better segmentation results. Our code will be available at https://github.com/YichiZhang98/SemiSAM.

\subsection{Baseline Approaches}

To make a comprehensive comparison of the improvements of SSL methods, we conduct experiments using only labeled set as the fully supervised (FS) baseline. Besides, an upper-bound performance is adopted using both labeled and unlabeled data with all annotations.
For specialist models, we conduct experiments on four representative semi-supervised frameworks including MT \citep{tarvainen2017mean}, UA-MT \citep{yu2019uncertainty},  DAN \citep{zhang2017deep}, and DTC \citep{luo2021semi} as introduced in Sec. \ref{method1}.
In addition, we also include various state-of-the-art semi-supervised segmentation methods with mixed strategies and different backbones, including UGMCL \citep{zhang2023uncertainty}, ACMT \citep{xu2023ambiguity}, and SDCL \citep{song2024sdcl}.
For generalist models, we make comparison experiments with foundation models for promptable segmentation \citep{SAM,MedSAM,SAM-Med3D}.
In addition, we also conduct experiments on SAM-based supervised segmentation framework \citep{autosam,SAMed,chen2024ma,cheng2024unleashing} for comparison.

\subsection{Comparison between Generalist and Specialist Models}

As shown in Table. \ref{Table_compare}, we compare the performance of generalist models for universal promptable segmentation and specialist models for training-based automatic segmentation on different segmentation tasks. For task-specific specialist models, manual annotations are required to train the model for each downstream segmentation task. The segmentation performance increases as the number of labeled training data increases. However, when only one or a few labeled images are available, the performance is far behind fully supervised performances (FS Upperbound). While SSL specialist models can utilize unlabeled data in conjunction with labeled data for training and improve the segmentation performance.
For generalist foundation models, it can be observed that an increase in the number of prompt points leads to improved segmentation performance. Additionally, by utilizing volumetric information, 3D generalist models like SAM-Med3D demonstrates superior performance over classic SAM with significantly fewer prompt points. Although these methods can achieve zero-shot segmentation without the need for annotated data for training, high manual efforts are still needed for generating prompt points of each test image to obtain acceptable performance. 
With the collaborative learning of specialist model and generalist model, SemiSAM+ significantly improve the performance of specialist model with the usage of pre-trained knowledge of generalist model.

\subsection{Ablation Analysis}
\label{ablation}

In this study, we performed comprehensive ablation studies on the LA dataset based on SAM-Med3D \citep{SAM-Med3D} as the generalist model and Mean Teacher \citep{tarvainen2017mean} as the specialist model to analyze the effectiveness of each design in the SemiSAM+ framework.
Firstly, we investigate the effectiveness of different prompting strategies. We conduct experiments on different strategies for generalist models, including directly utilizing the coarse segmentation as mask prompt (m) and generating point prompts based on the coarse segmentation (p).
As introduced in Sec. \ref{method3}, instead of directly enforcing the consistency of outputs between the specialist model and the generalist model, we utilize a confidence-aware regularization based on different prompting types (ur).
The experimental results are shown in Table \ref{Table_ablation}.
From the results, we can observe that in most scenarios, directly utilizing output of specialist model as mask prompt outperforms generating point prompts based on the mask, as the interactive prompting manner may cause error accumulation as the output mask is inaccurate when annotations are scarce, and the confidence-aware regularization can ensure more reliable guidance for the learning procedure.

\begin{figure*}[h]
	\includegraphics[width=18cm]{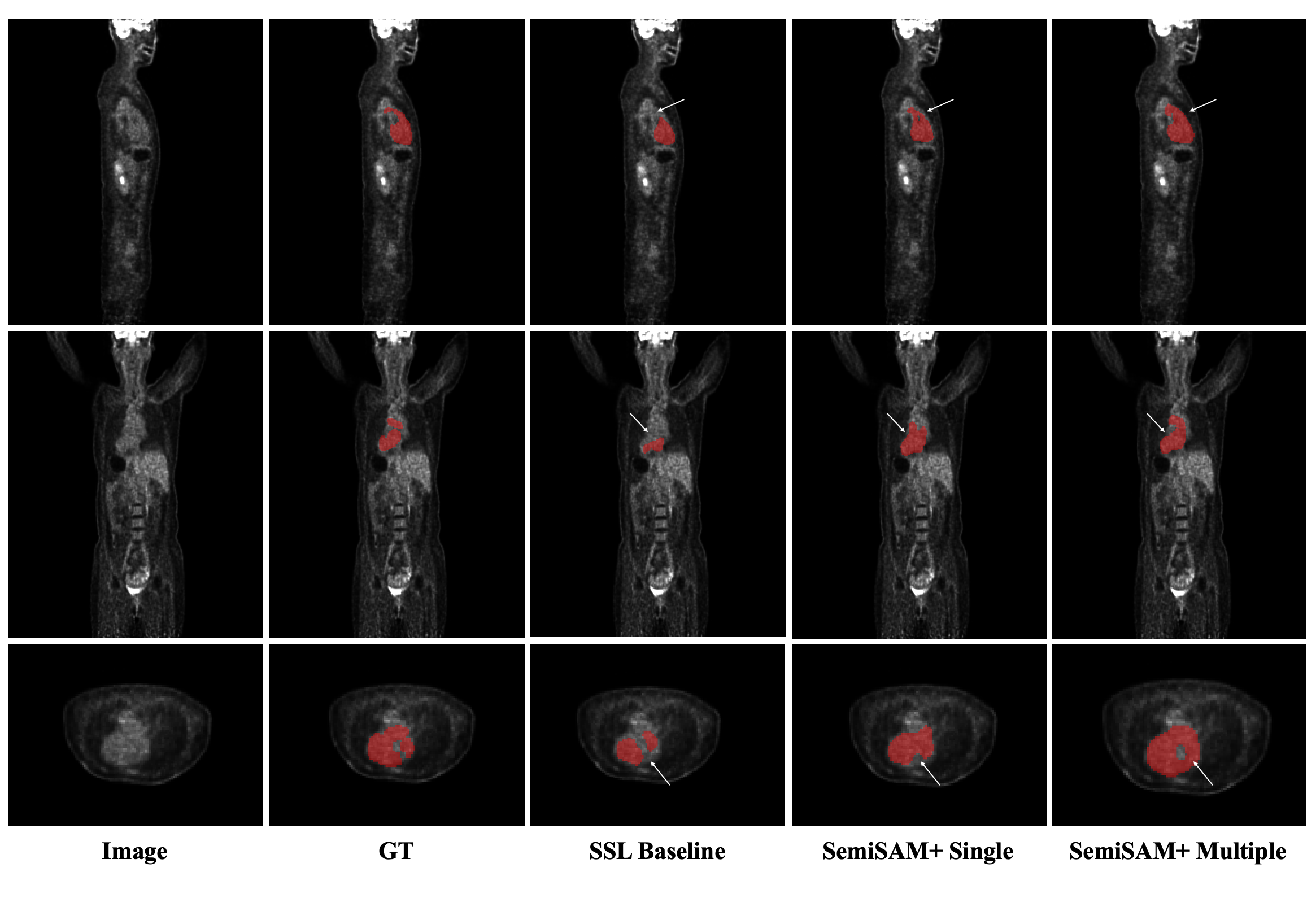}
	\caption{Visual comparison of segmentation results of SSL baseline (MT) and SemiSAM+ using single generalist model (SAM-Med3D) and multiple generalist models (SAM-Med3D and SegAnyPET) on whole heart segmentation from PET images.}
    \label{vis_pet}
\end{figure*}

\subsection{Compatibility to Different SSL Specialist Models}

With the plug-and-play nature of SemiSAM+, we conduct comprehensive experiments of SemiSAM+ upon four SSL specialist models with different learning strategies as introduced before and SAM-Med3D \citep{SAM-Med3D} as generalist model on left atrium segmentation and brain tumor segmentation tasks.
Table \ref{Table_specialist} presents the performance of these different implementations.
Despite the varying performance of different specialist models in differnet segmentation tasks and learning scenarios, we observe that SemiSAM+ can be efficiently adapted to different SSL strategies and significant improve performance in most situations. As shown in Fig. \ref{plot}, compared to the SSL baseline, SemiSAM+ yields excellent segmentation performance especially in extremely limited annotation scenarios.
Visualization of segmentation results in Fig. \ref{vis_labt} shows that SemiSAM+ predictions align more accurately with ground truth masks.

\subsection{Comparion with State-of-the-art Methods}

Table \ref{Table_sota} presents the performance of SemiSAM+ and more recent state-of-the-art semi-supervised approaches \citep{zhang2023uncertainty,xu2023ambiguity,song2024sdcl} on the left atrium segmentation task using different numbers of labeled images.
Specifically, we adopt SemiSAM+ upon SAM-Med3D \citep{SAM-Med3D} as the generalist model and Mean Teacher \citep{tarvainen2017mean} as the specialist model for comparison.
We also make comparison with SAM-based trainable segmentation framework including AutoSAM \citep{autosam}, SAMed \citep{SAMed}, MA-SAM \citep{chen2024ma} and H-SAM \citep{cheng2024unleashing}. Specifically, we implement another trainable adaption Auto-SAM-Med3D building upon 3D segmentation framework SAM-Med3D following the adaption of AutoSAM.
From the experimental results, we observe that building upon the vanilla MT, SemiSAM+ can significantly improve the performance and outperforms several cutting-edge semi-supervised medical image segmentation methods. For the most recent state-of-the-art SDCL, incorporating SemiSAM+ can still achieve considerable improvements.
Despite the comparable performance of SAM-based trainable segmentation frameworks, since the inference stage of SemiSAM+ is based on the specialist model, the parameter size maintains the same level as the backbone network (i.e. 3D U-Net), which is significantly smaller than that of these SAM-based models.

\subsection{Adaption to Multiple Generalist Models}

Despite the generalization ability of generalist models trained upon large-scale datasets covering multiple modalities and anatomical structures, their generalization performance to some specific task may not be good. For example, most existing generalist models \citep{SAM-Med3D,MedSAM} are not trained on PET images due to the lack of publicly available datasets. 
To evaluate the flexibility of SemiSAM+ to different and multiple generalist models, we conduct experiments using medical universal generalist model SAM-Med3D \citep{SAM-Med3D}, modality-specific generalist model SegAnyPET \citep{zhang2025seganypet}, and their combination upon the same Mean Teacher framework \citep{tarvainen2017mean} as specialist model for comparison. The quantitative performance and visual comparison of PET heart segmentation is shown in Table. \ref{Table_pet} and Fig. \ref{vis_pet}.
As shown in the table, using SegAnyPET achieves better performance than using SAM-Med3D as generalist model for training. This is likely due to the fact that the training data for SegAnyPET and downstream PET data share the same modality and have a smaller distribution gap. However, SAM-Med3D, which is trained on a larger-scale dataset that includes multiple different modalities, may have also learned representations that are useful for the segmentation task.
It can be observed that combining the guidance of these two generalist models together for training obtains better performance in most situations.
As a result, the design of SemiSAM+ also provides more flexibility for new updated versions of generalist models and the optional choice to learn from multiple generalist models with different focus and advantages for improved performance.

\section{Discussion and Conclusion}

Although deep learning-based automatic medical image segmentation has shown remarkable performance in many tasks, it faces a limitation in clinical application as it requires a large amount of annotated data for training. Semi-supervised learning focus on learning from a small amount of labeled data and large amount of unlabeled data, which has shown the potential to deal with this challenge.
However, these approaches still require a small subset of high-quality annotated data, struggling to achieve satisfactory outcomes when faced with extremely limited labeling budgets.
One possible reason is that the model cannot learn sufficient discriminative information with extremely limited supervision. As a result, the predictions are full of randomness and the improvement of exploiting unlabeled data is limited.

With the introduction of segmentation foundation models represented by the Segment Anything Model (SAM) \citep {SAM}, which have shown zero-shot segmentation capabilities for various new segmentation tasks with positional prompts.
Despite recent work \citep{SAM4MIS} revealing that SAM's unsatisfying performance for direct application to medical image segmentation tasks, it still offers new possibilities serving as a reliable pseudo-label generator to guide the learning procedure.
To this end, we propose SemiSAM+, a foundation model-driven annotation-efficient learning framework to efficiently learn from limited labeled data for medical image segmentation.
In contrast with existing model-centric semi-supervised methods for medical image segmentation, SemiSAM+ represents a new paradigm which focuses on utilizing pre-trained knowledge of generalist foundation models to assist in trainable specialist SSL model in a collaborative learning manner.
An advantage of SemiSAM+ is that it is a plug-and-play strategy that can be easily adapted to new trainable specialist models and generalist foundation models.

Due to the promptable manner of generalist foundation models, our experiments only focus on single-target segmentation.
However, the specialist model can be easily adapted to multi-target segmentation by changing the output class channels, and the generalist model can be adapted by prompting each target with corresponding class-level feedback.
Another limitation of current work is that the nature of point prompts hinders the application of the method into challenging non-compact targets with irregular shape. In the future, we aim to exploit more flexible prompt types of generalist models like scribble prompts \citep{wong2025scribbleprompt} and semantic text prompts \citep{du2023segvol} to further extend the usability.
We hope that this work will act as catalysts to promote future in-depth research in annotation-efficient medical image segmentation in the era of foundation models, ultimately benefiting clinical applicable healthcare practices.

\section*{Declaration of Competing Interests}

The authors declare that they have no known competing financial interests or personal relationships that could have appeared to influence the work reported in this paper.

\bibliographystyle{model2-names.bst}\biboptions{authoryear}
\bibliography{refs}

\begin{thebibliography}{59}
\expandafter\ifx\csname natexlab\endcsname\relax\def\natexlab#1{#1}\fi
\providecommand{\url}[1]{\texttt{#1}}
\providecommand{\href}[2]{#2}
\providecommand{\path}[1]{#1}
\providecommand{\DOIprefix}{doi:}
\providecommand{\ArXivprefix}{arXiv:}
\providecommand{\URLprefix}{URL: }
\providecommand{\Pubmedprefix}{pmid:}
\providecommand{\doi}[1]{\href{http://dx.doi.org/#1}{\path{#1}}}
\providecommand{\Pubmed}[1]{\href{pmid:#1}{\path{#1}}}
\providecommand{\bibinfo}[2]{#2}
\ifx\xfnm\relax \def\xfnm[#1]{\unskip,\space#1}\fi
\bibitem[{Adiga et~al.(2024)Adiga, Dolz and Lombaert}]{adiga2024anatomically}
\bibinfo{author}{Adiga, S.}, \bibinfo{author}{Dolz, J.}, \bibinfo{author}{Lombaert, H.}, \bibinfo{year}{2024}.
\newblock \bibinfo{title}{Anatomically-aware uncertainty for semi-supervised image segmentation}.
\newblock \bibinfo{journal}{Medical Image Analysis} \bibinfo{volume}{91}, \bibinfo{pages}{103011}.
\bibitem[{Ali et~al.(2024)Ali, Wu, Hu, Luo, Xu, Zheng, Jin, Yang and Yao}]{ali2024review}
\bibinfo{author}{Ali, M.}, \bibinfo{author}{Wu, T.}, \bibinfo{author}{Hu, H.}, \bibinfo{author}{Luo, Q.}, \bibinfo{author}{Xu, D.}, \bibinfo{author}{Zheng, W.}, \bibinfo{author}{Jin, N.}, \bibinfo{author}{Yang, C.}, \bibinfo{author}{Yao, J.}, \bibinfo{year}{2024}.
\newblock \bibinfo{title}{A review of the segment anything model (sam) for medical image analysis: Accomplishments and perspectives}.
\newblock \bibinfo{journal}{Computerized Medical Imaging and Graphics} , \bibinfo{pages}{102473}.
\bibitem[{Antonelli et~al.(2022)Antonelli, Reinke, Bakas, Farahani, Kopp-Schneider, Landman, Litjens, Menze, Ronneberger, Summers et~al.}]{MSD}
\bibinfo{author}{Antonelli, M.}, \bibinfo{author}{Reinke, A.}, \bibinfo{author}{Bakas, S.}, \bibinfo{author}{Farahani, K.}, \bibinfo{author}{Kopp-Schneider, A.}, \bibinfo{author}{Landman, B.A.}, \bibinfo{author}{Litjens, G.}, \bibinfo{author}{Menze, B.}, \bibinfo{author}{Ronneberger, O.}, \bibinfo{author}{Summers, R.M.}, et~al., \bibinfo{year}{2022}.
\newblock \bibinfo{title}{The medical segmentation decathlon}.
\newblock \bibinfo{journal}{Nature communications} \bibinfo{volume}{13}, \bibinfo{pages}{4128}.
\bibitem[{Awais et~al.(2023)Awais, Naseer, Khan, Anwer, Cholakkal, Shah, Yang and Khan}]{awais2023foundational}
\bibinfo{author}{Awais, M.}, \bibinfo{author}{Naseer, M.}, \bibinfo{author}{Khan, S.}, \bibinfo{author}{Anwer, R.M.}, \bibinfo{author}{Cholakkal, H.}, \bibinfo{author}{Shah, M.}, \bibinfo{author}{Yang, M.H.}, \bibinfo{author}{Khan, F.S.}, \bibinfo{year}{2023}.
\newblock \bibinfo{title}{Foundational models defining a new era in vision: A survey and outlook}.
\newblock \bibinfo{journal}{arXiv preprint arXiv:2307.13721} .
\bibitem[{Bakas(2020)}]{hdtd-5j88-20}
\bibinfo{author}{Bakas, S.S.}, \bibinfo{year}{2020}.
\newblock \bibinfo{title}{Brats miccai brain tumor dataset}.
\newblock \URLprefix \url{https://dx.doi.org/10.21227/hdtd-5j88}, \DOIprefix\doi{10.21227/hdtd-5j88}.
\bibitem[{Chen et~al.(2024)Chen, Miao, Wu, Zhong, Yan, Kim, Hu, Liu, Sun, Li et~al.}]{chen2024ma}
\bibinfo{author}{Chen, C.}, \bibinfo{author}{Miao, J.}, \bibinfo{author}{Wu, D.}, \bibinfo{author}{Zhong, A.}, \bibinfo{author}{Yan, Z.}, \bibinfo{author}{Kim, S.}, \bibinfo{author}{Hu, J.}, \bibinfo{author}{Liu, Z.}, \bibinfo{author}{Sun, L.}, \bibinfo{author}{Li, X.}, et~al., \bibinfo{year}{2024}.
\newblock \bibinfo{title}{Ma-sam: Modality-agnostic sam adaptation for 3d medical image segmentation}.
\newblock \bibinfo{journal}{Medical Image Analysis} \bibinfo{volume}{98}, \bibinfo{pages}{103310}.
\bibitem[{Chen et~al.(2023)Chen, Bai, Shen, Li, Yu and Wang}]{chen2023magicnet}
\bibinfo{author}{Chen, D.}, \bibinfo{author}{Bai, Y.}, \bibinfo{author}{Shen, W.}, \bibinfo{author}{Li, Q.}, \bibinfo{author}{Yu, L.}, \bibinfo{author}{Wang, Y.}, \bibinfo{year}{2023}.
\newblock \bibinfo{title}{Magicnet: Semi-supervised multi-organ segmentation via magic-cube partition and recovery}, in: \bibinfo{booktitle}{Proceedings of the IEEE/CVF Conference on Computer Vision and Pattern Recognition}, pp. \bibinfo{pages}{23869--23878}.
\bibitem[{Chen et~al.(2022)Chen, Sun, Wei, Wu and Ming}]{chen2022semi}
\bibinfo{author}{Chen, Q.Q.}, \bibinfo{author}{Sun, Z.H.}, \bibinfo{author}{Wei, C.F.}, \bibinfo{author}{Wu, E.Q.}, \bibinfo{author}{Ming, D.}, \bibinfo{year}{2022}.
\newblock \bibinfo{title}{Semi-supervised 3d medical image segmentation based on dual-task consistent joint learning and task-level regularization}.
\newblock \bibinfo{journal}{IEEE/ACM Transactions on Computational Biology and Bioinformatics} .
\bibitem[{Cheng et~al.(2023)Cheng, Ye, Deng, Chen, Li, Wang, Su, Huang, Chen, Jiang et~al.}]{SAM-Med2D}
\bibinfo{author}{Cheng, J.}, \bibinfo{author}{Ye, J.}, \bibinfo{author}{Deng, Z.}, \bibinfo{author}{Chen, J.}, \bibinfo{author}{Li, T.}, \bibinfo{author}{Wang, H.}, \bibinfo{author}{Su, Y.}, \bibinfo{author}{Huang, Z.}, \bibinfo{author}{Chen, J.}, \bibinfo{author}{Jiang, L.}, et~al., \bibinfo{year}{2023}.
\newblock \bibinfo{title}{Sam-med2d}.
\newblock \bibinfo{journal}{arXiv preprint arXiv:2308.16184} .
\bibitem[{Cheng et~al.(2024)Cheng, Wei, Zhu, Wang, Qu, Shao and Zhou}]{cheng2024unleashing}
\bibinfo{author}{Cheng, Z.}, \bibinfo{author}{Wei, Q.}, \bibinfo{author}{Zhu, H.}, \bibinfo{author}{Wang, Y.}, \bibinfo{author}{Qu, L.}, \bibinfo{author}{Shao, W.}, \bibinfo{author}{Zhou, Y.}, \bibinfo{year}{2024}.
\newblock \bibinfo{title}{Unleashing the potential of sam for medical adaptation via hierarchical decoding}, in: \bibinfo{booktitle}{Proceedings of the IEEE/CVF Conference on Computer Vision and Pattern Recognition}, pp. \bibinfo{pages}{3511--3522}.
\bibitem[{Cheplygina et~al.(2019)Cheplygina, de~Bruijne and Pluim}]{cheplygina2019not}
\bibinfo{author}{Cheplygina, V.}, \bibinfo{author}{de~Bruijne, M.}, \bibinfo{author}{Pluim, J.P.}, \bibinfo{year}{2019}.
\newblock \bibinfo{title}{Not-so-supervised: a survey of semi-supervised, multi-instance, and transfer learning in medical image analysis}.
\newblock \bibinfo{journal}{Medical image analysis} \bibinfo{volume}{54}, \bibinfo{pages}{280--296}.
\bibitem[{{\c{C}}i{\c{c}}ek et~al.(2016){\c{C}}i{\c{c}}ek, Abdulkadir, Lienkamp, Brox and Ronneberger}]{cciccek20163d}
\bibinfo{author}{{\c{C}}i{\c{c}}ek, {\"O}.}, \bibinfo{author}{Abdulkadir, A.}, \bibinfo{author}{Lienkamp, S.S.}, \bibinfo{author}{Brox, T.}, \bibinfo{author}{Ronneberger, O.}, \bibinfo{year}{2016}.
\newblock \bibinfo{title}{3d u-net: learning dense volumetric segmentation from sparse annotation}, in: \bibinfo{booktitle}{International conference on medical image computing and computer-assisted intervention}, \bibinfo{organization}{Springer}. pp. \bibinfo{pages}{424--432}.
\bibitem[{Du et~al.(2023)Du, Bai, Huang and Zhao}]{du2023segvol}
\bibinfo{author}{Du, Y.}, \bibinfo{author}{Bai, F.}, \bibinfo{author}{Huang, T.}, \bibinfo{author}{Zhao, B.}, \bibinfo{year}{2023}.
\newblock \bibinfo{title}{Segvol: Universal and interactive volumetric medical image segmentation}.
\newblock \bibinfo{journal}{arXiv preprint arXiv:2311.13385} .
\bibitem[{DuMont~Sch{\"u}tte et~al.(2021)DuMont~Sch{\"u}tte, Hetzel, Gatidis, Hepp, Dietz, Bauer and Schwab}]{dumont2021overcoming}
\bibinfo{author}{DuMont~Sch{\"u}tte, A.}, \bibinfo{author}{Hetzel, J.}, \bibinfo{author}{Gatidis, S.}, \bibinfo{author}{Hepp, T.}, \bibinfo{author}{Dietz, B.}, \bibinfo{author}{Bauer, S.}, \bibinfo{author}{Schwab, P.}, \bibinfo{year}{2021}.
\newblock \bibinfo{title}{Overcoming barriers to data sharing with medical image generation: a comprehensive evaluation}.
\newblock \bibinfo{journal}{NPJ digital medicine} \bibinfo{volume}{4}, \bibinfo{pages}{141}.
\bibitem[{Gu et~al.(2025)Gu, Colglazier, Dong, Zhang, Chen, Yildiz, Chen, Li, Yang, Willhite et~al.}]{gu2025segmentanybone}
\bibinfo{author}{Gu, H.}, \bibinfo{author}{Colglazier, R.}, \bibinfo{author}{Dong, H.}, \bibinfo{author}{Zhang, J.}, \bibinfo{author}{Chen, Y.}, \bibinfo{author}{Yildiz, Z.}, \bibinfo{author}{Chen, Y.}, \bibinfo{author}{Li, L.}, \bibinfo{author}{Yang, J.}, \bibinfo{author}{Willhite, J.}, et~al., \bibinfo{year}{2025}.
\newblock \bibinfo{title}{Segmentanybone: A universal model that segments any bone at any location on mri}.
\newblock \bibinfo{journal}{Medical Image Analysis} , \bibinfo{pages}{103469}.
\bibitem[{Huang et~al.(2022)Huang, Chen, Lin, Cai, Zhang, Iwamoto, Han, Furukawa, Kanasaki, Chen et~al.}]{huang2022mtl}
\bibinfo{author}{Huang, H.}, \bibinfo{author}{Chen, Q.}, \bibinfo{author}{Lin, L.}, \bibinfo{author}{Cai, M.}, \bibinfo{author}{Zhang, Q.W.}, \bibinfo{author}{Iwamoto, Y.}, \bibinfo{author}{Han, X.}, \bibinfo{author}{Furukawa, A.}, \bibinfo{author}{Kanasaki, S.}, \bibinfo{author}{Chen, Y.W.}, et~al., \bibinfo{year}{2022}.
\newblock \bibinfo{title}{Mtl-abs3net: Atlas-based semi-supervised organ segmentation network with multi-task learning for medical images}.
\newblock \bibinfo{journal}{IEEE Journal of Biomedical and Health Informatics} .
\bibitem[{Jiao et~al.(2023)Jiao, Zhang, Ding, Xue, Zhang, Cai and Jin}]{SemiSurvey}
\bibinfo{author}{Jiao, R.}, \bibinfo{author}{Zhang, Y.}, \bibinfo{author}{Ding, L.}, \bibinfo{author}{Xue, B.}, \bibinfo{author}{Zhang, J.}, \bibinfo{author}{Cai, R.}, \bibinfo{author}{Jin, C.}, \bibinfo{year}{2023}.
\newblock \bibinfo{title}{Learning with limited annotations: A survey on deep semi-supervised learning for medical image segmentation}.
\newblock \bibinfo{journal}{Computers in Biology and Medicine} .
\bibitem[{Kirillov et~al.(2023)Kirillov, Mintun, Ravi, Mao, Rolland, Gustafson, Xiao, Whitehead, Berg, Lo et~al.}]{SAM}
\bibinfo{author}{Kirillov, A.}, \bibinfo{author}{Mintun, E.}, \bibinfo{author}{Ravi, N.}, \bibinfo{author}{Mao, H.}, \bibinfo{author}{Rolland, C.}, \bibinfo{author}{Gustafson, L.}, \bibinfo{author}{Xiao, T.}, \bibinfo{author}{Whitehead, S.}, \bibinfo{author}{Berg, A.C.}, \bibinfo{author}{Lo, W.Y.}, et~al., \bibinfo{year}{2023}.
\newblock \bibinfo{title}{Segment anything}.
\newblock \bibinfo{journal}{arXiv preprint arXiv:2304.02643} .
\bibitem[{Lalande et~al.(2022)Lalande, Chen, Pommier, Decourselle, Qayyum, Salomon, Ginhac, Skandarani, Boucher, Brahim et~al.}]{lalande2021deep}
\bibinfo{author}{Lalande, A.}, \bibinfo{author}{Chen, Z.}, \bibinfo{author}{Pommier, T.}, \bibinfo{author}{Decourselle, T.}, \bibinfo{author}{Qayyum, A.}, \bibinfo{author}{Salomon, M.}, \bibinfo{author}{Ginhac, D.}, \bibinfo{author}{Skandarani, Y.}, \bibinfo{author}{Boucher, A.}, \bibinfo{author}{Brahim, K.}, et~al., \bibinfo{year}{2022}.
\newblock \bibinfo{title}{Deep learning methods for automatic evaluation of delayed enhancement-mri. the results of the emidec challenge}.
\newblock \bibinfo{journal}{Medical Image Analysis} \bibinfo{volume}{79}, \bibinfo{pages}{102428}.
\bibitem[{Lee et~al.(2013)}]{lee2013pseudo}
\bibinfo{author}{Lee, D.H.}, et~al., \bibinfo{year}{2013}.
\newblock \bibinfo{title}{Pseudo-label: The simple and efficient semi-supervised learning method for deep neural networks}, in: \bibinfo{booktitle}{Workshop on challenges in representation learning, ICML}, \bibinfo{organization}{Atlanta}. p. \bibinfo{pages}{896}.
\bibitem[{Li et~al.(2023)Li, Zhang, Sun, Zou, Liu, Yang, Li, Zhang and Gao}]{semanticSAM}
\bibinfo{author}{Li, F.}, \bibinfo{author}{Zhang, H.}, \bibinfo{author}{Sun, P.}, \bibinfo{author}{Zou, X.}, \bibinfo{author}{Liu, S.}, \bibinfo{author}{Yang, J.}, \bibinfo{author}{Li, C.}, \bibinfo{author}{Zhang, L.}, \bibinfo{author}{Gao, J.}, \bibinfo{year}{2023}.
\newblock \bibinfo{title}{Semantic-sam: Segment and recognize anything at any granularity}.
\newblock \bibinfo{journal}{arXiv preprint arXiv:2307.04767} .
\bibitem[{Li et~al.(2020)Li, Zhang and He}]{li2020shape}
\bibinfo{author}{Li, S.}, \bibinfo{author}{Zhang, C.}, \bibinfo{author}{He, X.}, \bibinfo{year}{2020}.
\newblock \bibinfo{title}{Shape-aware semi-supervised 3d semantic segmentation for medical images}, in: \bibinfo{booktitle}{International Conference on Medical Image Computing and Computer-Assisted Intervention}, \bibinfo{organization}{Springer}. pp. \bibinfo{pages}{552--561}.
\bibitem[{Luo et~al.(2021a)Luo, Chen, Song and Wang}]{luo2021semi}
\bibinfo{author}{Luo, X.}, \bibinfo{author}{Chen, J.}, \bibinfo{author}{Song, T.}, \bibinfo{author}{Wang, G.}, \bibinfo{year}{2021}a.
\newblock \bibinfo{title}{Semi-supervised medical image segmentation through dual-task consistency}, in: \bibinfo{booktitle}{Proceedings of the AAAI Conference on Artificial Intelligence}, pp. \bibinfo{pages}{8801--8809}.
\bibitem[{Luo et~al.(2021b)Luo, Hu, Song, Wang and Zhang}]{luo2021crossteaching}
\bibinfo{author}{Luo, X.}, \bibinfo{author}{Hu, M.}, \bibinfo{author}{Song, T.}, \bibinfo{author}{Wang, G.}, \bibinfo{author}{Zhang, S.}, \bibinfo{year}{2021}b.
\newblock \bibinfo{title}{Semi-supervised medical image segmentation via cross teaching between cnn and transformer}.
\newblock \bibinfo{journal}{arXiv preprint arXiv:2112.04894} .
\bibitem[{Luo et~al.(2022)Luo, Wang, Liao, Chen, Song, Chen, Zhang, Metaxas and Zhang}]{luo2022semi}
\bibinfo{author}{Luo, X.}, \bibinfo{author}{Wang, G.}, \bibinfo{author}{Liao, W.}, \bibinfo{author}{Chen, J.}, \bibinfo{author}{Song, T.}, \bibinfo{author}{Chen, Y.}, \bibinfo{author}{Zhang, S.}, \bibinfo{author}{Metaxas, D.N.}, \bibinfo{author}{Zhang, S.}, \bibinfo{year}{2022}.
\newblock \bibinfo{title}{Semi-supervised medical image segmentation via uncertainty rectified pyramid consistency}.
\newblock \bibinfo{journal}{Medical Image Analysis} , \bibinfo{pages}{102517}.
\bibitem[{Ma and Wang(2024)}]{ma2024semi}
\bibinfo{author}{Ma, C.}, \bibinfo{author}{Wang, Z.}, \bibinfo{year}{2024}.
\newblock \bibinfo{title}{Semi-mamba-unet: Pixel-level contrastive and cross-supervised visual mamba-based unet for semi-supervised medical image segmentation}.
\newblock \bibinfo{journal}{Knowledge-Based Systems} \bibinfo{volume}{300}, \bibinfo{pages}{112203}.
\bibitem[{Ma et~al.(2024)Ma, He, Li, Han, You and Wang}]{MedSAM}
\bibinfo{author}{Ma, J.}, \bibinfo{author}{He, Y.}, \bibinfo{author}{Li, F.}, \bibinfo{author}{Han, L.}, \bibinfo{author}{You, C.}, \bibinfo{author}{Wang, B.}, \bibinfo{year}{2024}.
\newblock \bibinfo{title}{Segment anything in medical images}.
\newblock \bibinfo{journal}{Nature Communications} \bibinfo{volume}{15}, \bibinfo{pages}{1--9}.
\bibitem[{Ma et~al.(2022)Ma, Zhang, Gu, Zhu, Ge, Zhang, An, Wang, Wang, Liu, Cao, Zhang, Liu, Wang, Li, He and Yang}]{AbdomenCT-1K}
\bibinfo{author}{Ma, J.}, \bibinfo{author}{Zhang, Y.}, \bibinfo{author}{Gu, S.}, \bibinfo{author}{Zhu, C.}, \bibinfo{author}{Ge, C.}, \bibinfo{author}{Zhang, Y.}, \bibinfo{author}{An, X.}, \bibinfo{author}{Wang, C.}, \bibinfo{author}{Wang, Q.}, \bibinfo{author}{Liu, X.}, \bibinfo{author}{Cao, S.}, \bibinfo{author}{Zhang, Q.}, \bibinfo{author}{Liu, S.}, \bibinfo{author}{Wang, Y.}, \bibinfo{author}{Li, Y.}, \bibinfo{author}{He, J.}, \bibinfo{author}{Yang, X.}, \bibinfo{year}{2022}.
\newblock \bibinfo{title}{Abdomenct-1k: Is abdominal organ segmentation a solved problem?}
\newblock \bibinfo{journal}{IEEE Transactions on Pattern Analysis and Machine Intelligence} \bibinfo{volume}{44}, \bibinfo{pages}{6695--6714}.
\bibitem[{Mazurowski et~al.(2023)Mazurowski, Dong, Gu, Yang, Konz and Zhang}]{SAM-Empirical}
\bibinfo{author}{Mazurowski, M.A.}, \bibinfo{author}{Dong, H.}, \bibinfo{author}{Gu, H.}, \bibinfo{author}{Yang, J.}, \bibinfo{author}{Konz, N.}, \bibinfo{author}{Zhang, Y.}, \bibinfo{year}{2023}.
\newblock \bibinfo{title}{Segment anything model for medical image analysis: an experimental study}.
\newblock \bibinfo{journal}{Medical Image Analysis} \bibinfo{volume}{89}, \bibinfo{pages}{102918}.
\bibitem[{Moor et~al.(2023)Moor, Banerjee, Abad, Krumholz, Leskovec, Topol and Rajpurkar}]{moor2023foundation}
\bibinfo{author}{Moor, M.}, \bibinfo{author}{Banerjee, O.}, \bibinfo{author}{Abad, Z.F.H.}, \bibinfo{author}{Krumholz, H.M.}, \bibinfo{author}{Leskovec, J.}, \bibinfo{author}{Topol, E.J.}, \bibinfo{author}{Rajpurkar, P.}, \bibinfo{year}{2023}.
\newblock \bibinfo{title}{Foundation models for generalist medical artificial intelligence}.
\newblock \bibinfo{journal}{Nature} \bibinfo{volume}{616}, \bibinfo{pages}{259--265}.
\bibitem[{Seibold et~al.(2022)Seibold, Rei{\ss}, Kleesiek and Stiefelhagen}]{seibold2021reference}
\bibinfo{author}{Seibold, C.M.}, \bibinfo{author}{Rei{\ss}, S.}, \bibinfo{author}{Kleesiek, J.}, \bibinfo{author}{Stiefelhagen, R.}, \bibinfo{year}{2022}.
\newblock \bibinfo{title}{Reference-guided pseudo-label generation for medical semantic segmentation} \bibinfo{volume}{36}, \bibinfo{pages}{2171--2179}.
\bibitem[{Shaharabany et~al.(2023)Shaharabany, Dahan, Giryes and Wolf}]{autosam}
\bibinfo{author}{Shaharabany, T.}, \bibinfo{author}{Dahan, A.}, \bibinfo{author}{Giryes, R.}, \bibinfo{author}{Wolf, L.}, \bibinfo{year}{2023}.
\newblock \bibinfo{title}{Autosam: Adapting sam to medical images by overloading the prompt encoder}.
\newblock \bibinfo{journal}{arXiv preprint arXiv:2306.06370} .
\bibitem[{Shi et~al.(2024)Shi, Ma, Yang, Wang and Zhang}]{shi2024beyond}
\bibinfo{author}{Shi, Y.}, \bibinfo{author}{Ma, J.}, \bibinfo{author}{Yang, J.}, \bibinfo{author}{Wang, S.}, \bibinfo{author}{Zhang, Y.}, \bibinfo{year}{2024}.
\newblock \bibinfo{title}{Beyond pixel-wise supervision for medical image segmentation: From traditional models to foundation models}.
\newblock \bibinfo{journal}{arXiv preprint arXiv:2404.13239} .
\bibitem[{Song and Wang(2024)}]{song2024sdcl}
\bibinfo{author}{Song, B.}, \bibinfo{author}{Wang, Q.}, \bibinfo{year}{2024}.
\newblock \bibinfo{title}{Sdcl: Students discrepancy-informed correction learning for semi-supervised medical image segmentation}, in: \bibinfo{booktitle}{International Conference on Medical Image Computing and Computer-Assisted Intervention}, \bibinfo{organization}{Springer}. pp. \bibinfo{pages}{567--577}.
\bibitem[{Tajbakhsh et~al.(2020)Tajbakhsh, Jeyaseelan, Li, Chiang, Wu and Ding}]{tajbakhsh2020embracing}
\bibinfo{author}{Tajbakhsh, N.}, \bibinfo{author}{Jeyaseelan, L.}, \bibinfo{author}{Li, Q.}, \bibinfo{author}{Chiang, J.N.}, \bibinfo{author}{Wu, Z.}, \bibinfo{author}{Ding, X.}, \bibinfo{year}{2020}.
\newblock \bibinfo{title}{Embracing imperfect datasets: A review of deep learning solutions for medical image segmentation}.
\newblock \bibinfo{journal}{Medical Image Analysis} \bibinfo{volume}{63}, \bibinfo{pages}{101693}.
\bibitem[{Tarvainen and Valpola(2017)}]{tarvainen2017mean}
\bibinfo{author}{Tarvainen, A.}, \bibinfo{author}{Valpola, H.}, \bibinfo{year}{2017}.
\newblock \bibinfo{title}{Mean teachers are better role models: Weight-averaged consistency targets improve semi-supervised deep learning results}.
\bibitem[{Wang et~al.(2023)Wang, Guo, Ye, Deng, Cheng, Li, Chen, Su, Huang, Shen, Fu et~al.}]{SAM-Med3D}
\bibinfo{author}{Wang, H.}, \bibinfo{author}{Guo, S.}, \bibinfo{author}{Ye, J.}, \bibinfo{author}{Deng, Z.}, \bibinfo{author}{Cheng, J.}, \bibinfo{author}{Li, T.}, \bibinfo{author}{Chen, J.}, \bibinfo{author}{Su, Y.}, \bibinfo{author}{Huang, Z.}, \bibinfo{author}{Shen, Y.}, \bibinfo{author}{Fu, B.}, et~al., \bibinfo{year}{2023}.
\newblock \bibinfo{title}{Sam-med3d}.
\newblock \bibinfo{journal}{arXiv preprint arXiv:2310.15161} .
\bibitem[{Wang et~al.(2022)Wang, Yuan, Guo, Huang, Cui, Xia, Wang, Bai and Chen}]{wang2022ssa}
\bibinfo{author}{Wang, X.}, \bibinfo{author}{Yuan, Y.}, \bibinfo{author}{Guo, D.}, \bibinfo{author}{Huang, X.}, \bibinfo{author}{Cui, Y.}, \bibinfo{author}{Xia, M.}, \bibinfo{author}{Wang, Z.}, \bibinfo{author}{Bai, C.}, \bibinfo{author}{Chen, S.}, \bibinfo{year}{2022}.
\newblock \bibinfo{title}{Ssa-net: Spatial self-attention network for covid-19 pneumonia infection segmentation with semi-supervised few-shot learning}.
\newblock \bibinfo{journal}{Medical Image Analysis} \bibinfo{volume}{79}, \bibinfo{pages}{102459}.
\bibitem[{Willemink et~al.(2022)Willemink, Roth and Sandfort}]{willemink2022toward}
\bibinfo{author}{Willemink, M.J.}, \bibinfo{author}{Roth, H.R.}, \bibinfo{author}{Sandfort, V.}, \bibinfo{year}{2022}.
\newblock \bibinfo{title}{Toward foundational deep learning models for medical imaging in the new era of transformer networks.}
\newblock \bibinfo{journal}{Radiology. Artificial intelligence} \bibinfo{volume}{46}, \bibinfo{pages}{210284}.
\bibitem[{Wong et~al.(2025)Wong, Rakic, Guttag and Dalca}]{wong2025scribbleprompt}
\bibinfo{author}{Wong, H.E.}, \bibinfo{author}{Rakic, M.}, \bibinfo{author}{Guttag, J.}, \bibinfo{author}{Dalca, A.V.}, \bibinfo{year}{2025}.
\newblock \bibinfo{title}{Scribbleprompt: fast and flexible interactive segmentation for any biomedical image}, in: \bibinfo{booktitle}{European Conference on Computer Vision}, \bibinfo{organization}{Springer}. pp. \bibinfo{pages}{207--229}.
\bibitem[{Xie et~al.(2023)Xie, Fu, Zheng, Zheng, Sham and Wang}]{xie2023adversarial}
\bibinfo{author}{Xie, H.}, \bibinfo{author}{Fu, C.}, \bibinfo{author}{Zheng, X.}, \bibinfo{author}{Zheng, Y.}, \bibinfo{author}{Sham, C.W.}, \bibinfo{author}{Wang, X.}, \bibinfo{year}{2023}.
\newblock \bibinfo{title}{Adversarial co-training for semantic segmentation over medical images}.
\newblock \bibinfo{journal}{Computers in biology and medicine} \bibinfo{volume}{157}, \bibinfo{pages}{106736}.
\bibitem[{Xiong et~al.(2021)Xiong, Xia, Hu, Huang, Bian, Zheng, Vesal, Ravikumar, Maier, Yang et~al.}]{xiong2021global}
\bibinfo{author}{Xiong, Z.}, \bibinfo{author}{Xia, Q.}, \bibinfo{author}{Hu, Z.}, \bibinfo{author}{Huang, N.}, \bibinfo{author}{Bian, C.}, \bibinfo{author}{Zheng, Y.}, \bibinfo{author}{Vesal, S.}, \bibinfo{author}{Ravikumar, N.}, \bibinfo{author}{Maier, A.}, \bibinfo{author}{Yang, X.}, et~al., \bibinfo{year}{2021}.
\newblock \bibinfo{title}{A global benchmark of algorithms for segmenting the left atrium from late gadolinium-enhanced cardiac magnetic resonance imaging}.
\newblock \bibinfo{journal}{Medical Image Analysis} \bibinfo{volume}{67}, \bibinfo{pages}{101832}.
\bibitem[{Xu et~al.(2023)Xu, Wang, Lu, Luo, Yan, Zheng and Tong}]{xu2023ambiguity}
\bibinfo{author}{Xu, Z.}, \bibinfo{author}{Wang, Y.}, \bibinfo{author}{Lu, D.}, \bibinfo{author}{Luo, X.}, \bibinfo{author}{Yan, J.}, \bibinfo{author}{Zheng, Y.}, \bibinfo{author}{Tong, R.K.y.}, \bibinfo{year}{2023}.
\newblock \bibinfo{title}{Ambiguity-selective consistency regularization for mean-teacher semi-supervised medical image segmentation}.
\newblock \bibinfo{journal}{Medical Image Analysis} \bibinfo{volume}{88}, \bibinfo{pages}{102880}.
\bibitem[{Yang et~al.(2023)Yang, Gao, Li, Gao, Wang and Zheng}]{trackanything}
\bibinfo{author}{Yang, J.}, \bibinfo{author}{Gao, M.}, \bibinfo{author}{Li, Z.}, \bibinfo{author}{Gao, S.}, \bibinfo{author}{Wang, F.}, \bibinfo{author}{Zheng, F.}, \bibinfo{year}{2023}.
\newblock \bibinfo{title}{Track anything: Segment anything meets videos}.
\newblock \bibinfo{journal}{arXiv preprint arXiv:2304.11968} .
\bibitem[{Yu et~al.(2019)Yu, Wang, Li, Fu and Heng}]{yu2019uncertainty}
\bibinfo{author}{Yu, L.}, \bibinfo{author}{Wang, S.}, \bibinfo{author}{Li, X.}, \bibinfo{author}{Fu, C.W.}, \bibinfo{author}{Heng, P.A.}, \bibinfo{year}{2019}.
\newblock \bibinfo{title}{Uncertainty-aware self-ensembling model for semi-supervised 3d left atrium segmentation}, in: \bibinfo{booktitle}{Medical Image Computing and Computer Assisted Intervention--MICCAI 2019: 22nd International Conference, Shenzhen, China, October 13--17, 2019, Proceedings, Part II 22}, \bibinfo{organization}{Springer}. pp. \bibinfo{pages}{605--613}.
\bibitem[{Zeineldin et~al.(2020)Zeineldin, Karar, Coburger, Wirtz and Burgert}]{zeineldin2020deepseg}
\bibinfo{author}{Zeineldin, R.A.}, \bibinfo{author}{Karar, M.E.}, \bibinfo{author}{Coburger, J.}, \bibinfo{author}{Wirtz, C.R.}, \bibinfo{author}{Burgert, O.}, \bibinfo{year}{2020}.
\newblock \bibinfo{title}{Deepseg: deep neural network framework for automatic brain tumor segmentation using magnetic resonance flair images}.
\newblock \bibinfo{journal}{International journal of computer assisted radiology and surgery} \bibinfo{volume}{15}, \bibinfo{pages}{909--920}.
\bibitem[{Zhang and Liu(2023)}]{SAMed}
\bibinfo{author}{Zhang, K.}, \bibinfo{author}{Liu, D.}, \bibinfo{year}{2023}.
\newblock \bibinfo{title}{Customized segment anything model for medical image segmentation}.
\newblock \bibinfo{journal}{arXiv preprint arXiv:2304.13785} .
\bibitem[{Zhang and Metaxas(2024)}]{zhang2024challenges}
\bibinfo{author}{Zhang, S.}, \bibinfo{author}{Metaxas, D.}, \bibinfo{year}{2024}.
\newblock \bibinfo{title}{On the challenges and perspectives of foundation models for medical image analysis}.
\newblock \bibinfo{journal}{Medical image analysis} \bibinfo{volume}{91}, \bibinfo{pages}{102996}.
\bibitem[{Zhang et~al.(2023)Zhang, Jiao, Liao, Li and Zhang}]{zhang2023uncertainty}
\bibinfo{author}{Zhang, Y.}, \bibinfo{author}{Jiao, R.}, \bibinfo{author}{Liao, Q.}, \bibinfo{author}{Li, D.}, \bibinfo{author}{Zhang, J.}, \bibinfo{year}{2023}.
\newblock \bibinfo{title}{Uncertainty-guided mutual consistency learning for semi-supervised medical image segmentation}.
\newblock \bibinfo{journal}{Artificial Intelligence in Medicine} \bibinfo{volume}{138}, \bibinfo{pages}{102476}.
\bibitem[{Zhang et~al.(2022)Zhang, Liao, Ding and Zhang}]{zhang2022bridging}
\bibinfo{author}{Zhang, Y.}, \bibinfo{author}{Liao, Q.}, \bibinfo{author}{Ding, L.}, \bibinfo{author}{Zhang, J.}, \bibinfo{year}{2022}.
\newblock \bibinfo{title}{Bridging 2d and 3d segmentation networks for computation-efficient volumetric medical image segmentation: An empirical study of 2.5 d solutions}, \bibinfo{publisher}{Elsevier}. p. \bibinfo{pages}{102088}.
\bibitem[{Zhang et~al.(2021)Zhang, Liao, Yuan, Zhu, Xing and Zhang}]{zhang2021exploiting}
\bibinfo{author}{Zhang, Y.}, \bibinfo{author}{Liao, Q.}, \bibinfo{author}{Yuan, L.}, \bibinfo{author}{Zhu, H.}, \bibinfo{author}{Xing, J.}, \bibinfo{author}{Zhang, J.}, \bibinfo{year}{2021}.
\newblock \bibinfo{title}{Exploiting shared knowledge from non-covid lesions for annotation-efficient covid-19 ct lung infection segmentation}.
\newblock \bibinfo{journal}{IEEE journal of biomedical and health informatics} \bibinfo{volume}{25}, \bibinfo{pages}{4152--4162}.
\bibitem[{Zhang and Shen(2024)}]{SAM2-MIS}
\bibinfo{author}{Zhang, Y.}, \bibinfo{author}{Shen, Z.}, \bibinfo{year}{2024}.
\newblock \bibinfo{title}{Unleashing the potential of sam2 for biomedical images and videos: A survey}.
\newblock \bibinfo{journal}{arXiv preprint arXiv:2408.12889} .
\bibitem[{Zhang et~al.(2024a)Zhang, Shen and Jiao}]{SAM4MIS}
\bibinfo{author}{Zhang, Y.}, \bibinfo{author}{Shen, Z.}, \bibinfo{author}{Jiao, R.}, \bibinfo{year}{2024}a.
\newblock \bibinfo{title}{Segment anything model for medical image segmentation: Current applications and future directions}.
\newblock \bibinfo{journal}{Computers in Biology and Medicine} \bibinfo{volume}{171}, \bibinfo{pages}{108238}.
\bibitem[{Zhang et~al.(2024b)Zhang, Wang, Pan, Jiang, Ge, Guo, Jiang, Lu, Zhang, Liu et~al.}]{zhang2024nasalseg}
\bibinfo{author}{Zhang, Y.}, \bibinfo{author}{Wang, J.}, \bibinfo{author}{Pan, T.}, \bibinfo{author}{Jiang, Q.}, \bibinfo{author}{Ge, J.}, \bibinfo{author}{Guo, X.}, \bibinfo{author}{Jiang, C.}, \bibinfo{author}{Lu, J.}, \bibinfo{author}{Zhang, J.}, \bibinfo{author}{Liu, X.}, et~al., \bibinfo{year}{2024}b.
\newblock \bibinfo{title}{Nasalseg: A dataset for automatic segmentation of nasal cavity and paranasal sinuses from 3d ct images}.
\newblock \bibinfo{journal}{Scientific Data} \bibinfo{volume}{11}, \bibinfo{pages}{1--5}.
\bibitem[{Zhang et~al.(2025)Zhang, Xue, Zhang, Li, Liu, Jiang, Cheng and Qi}]{zhang2025seganypet}
\bibinfo{author}{Zhang, Y.}, \bibinfo{author}{Xue, L.}, \bibinfo{author}{Zhang, W.}, \bibinfo{author}{Li, L.}, \bibinfo{author}{Liu, Y.}, \bibinfo{author}{Jiang, C.}, \bibinfo{author}{Cheng, Y.}, \bibinfo{author}{Qi, Y.}, \bibinfo{year}{2025}.
\newblock \bibinfo{title}{Seganypet: Universal promptable segmentation from positron emission tomography images}.
\newblock \bibinfo{journal}{arXiv preprint arXiv:2502.14351} .
\bibitem[{Zhang et~al.(2024c)Zhang, Yang, Liu, Cheng and Qi}]{zhang2024semisam}
\bibinfo{author}{Zhang, Y.}, \bibinfo{author}{Yang, J.}, \bibinfo{author}{Liu, Y.}, \bibinfo{author}{Cheng, Y.}, \bibinfo{author}{Qi, Y.}, \bibinfo{year}{2024}c.
\newblock \bibinfo{title}{Semisam: Enhancing semi-supervised medical image segmentation via sam-assisted consistency regularization}, in: \bibinfo{booktitle}{2024 IEEE International Conference on Bioinformatics and Biomedicine (BIBM)}, \bibinfo{organization}{IEEE}. pp. \bibinfo{pages}{3982--3986}.
\bibitem[{Zhang et~al.(2017)Zhang, Yang, Chen, Fredericksen, Hughes and Chen}]{zhang2017deep}
\bibinfo{author}{Zhang, Y.}, \bibinfo{author}{Yang, L.}, \bibinfo{author}{Chen, J.}, \bibinfo{author}{Fredericksen, M.}, \bibinfo{author}{Hughes, D.P.}, \bibinfo{author}{Chen, D.Z.}, \bibinfo{year}{2017}.
\newblock \bibinfo{title}{Deep adversarial networks for biomedical image segmentation utilizing unannotated images}, in: \bibinfo{booktitle}{International Conference on Medical Image Computing and Computer-Assisted Intervention}, \bibinfo{organization}{Springer}. pp. \bibinfo{pages}{408--416}.
\bibitem[{Zhang and Zhang(2021)}]{zhang2021dual}
\bibinfo{author}{Zhang, Y.}, \bibinfo{author}{Zhang, J.}, \bibinfo{year}{2021}.
\newblock \bibinfo{title}{Dual-task mutual learning for semi-supervised medical image segmentation}, in: \bibinfo{booktitle}{Chinese Conference on Pattern Recognition and Computer Vision (PRCV)}, \bibinfo{organization}{Springer}. pp. \bibinfo{pages}{548--559}.
\bibitem[{Zhao et~al.(2023)Zhao, Shen, Chen, Wang, Zhuang, Wang and Zhang}]{zhao2023one}
\bibinfo{author}{Zhao, X.}, \bibinfo{author}{Shen, Z.}, \bibinfo{author}{Chen, D.}, \bibinfo{author}{Wang, S.}, \bibinfo{author}{Zhuang, Z.}, \bibinfo{author}{Wang, Q.}, \bibinfo{author}{Zhang, L.}, \bibinfo{year}{2023}.
\newblock \bibinfo{title}{One-shot traumatic brain segmentation with adversarial training and uncertainty rectification}, in: \bibinfo{booktitle}{International Conference on Medical Image Computing and Computer-Assisted Intervention}, \bibinfo{organization}{Springer}. pp. \bibinfo{pages}{120--129}.

\end{thebibliography}

\end{document}